# Luminous efficiency based on FRIPON meteors

Esther Drolshagen[1*], Theresa Ott[1*], Detlef Koschny[2,3], Gerhard Drolshagen[1], Jeremie Vaubaillon[4], Francois Colas[4], Josep Maria Trigo-Rodriguez[5,6,7], Brigitte Zanda[8,4,9], Sylvain Bouley[10,4,9], Simon Jeanne[4,9], Adrien Malgoyre[11,9], Mirel Birlan[4,9], Pierre Vernazza[12,9], Daniele Gardiol[13], Dan Alin Nedelcu[14,15], Jim Rowe[16], Mathieu Forcier[17,18], Eloy Peña Asensio[6], Herve Lamy[19,20], Ludovic Ferrière[21,22], Dario Barghini[13,23], Albino Carbognani[24], Mario Di Martino[13], Stefania Rasetti[13], Giovanni Battista Valsecchi[25,26], Cosimo Antonio Volpicelli[13], Matteo Di Carlo[13], Cristina Knapic[13], Giovanni Pratesi[13,28], Walter Riva[33], Giovanna M. Stirpe[13], Sonia Zorba[13], Olivier Hernandez[17,18], Emmanuel Jehin[19,27], Marc Jobin[17,18], Ashley King[16,29], Agustin Sanchez-Lavega[30,31], Andrea Toni[32,2], and Björn Poppe[1]

*[1]University of Oldenburg, Division for Medical Radiation Physics and Space Environment, Germany.*
*[2] European Space Agency, OPS-SP, Keplerlaan 1, 2201 AZ Noordwijk, The Netherlands.*
*[3]Chair of Astronautics, TU Munich, Germany.*
*[4]IMCCE, Observatoire de Paris, PSL Research University, CNRS UMR 8028, Sorbonne Université, France.*
*[5]SPMN (SPanish Meteor Network), FRIPON, Spain.*
*[6]Institute of Space Sciences (CSIC), Campus UAB, Facultat de Ciències, 08193 Bellaterra, Barcelona, Catalonia, Spain.*
*[7]Institut d'Estudis Espacials de Catalunya (IEEC), 08034 Barcelona, Catalonia, Spain.*
*[8]Institut de Minéralogie, Physique des Matériaux et Cosmochimie (IMPMC), Muséum National d'Histoire Naturelle, CNRS UMR 7590, Sorbonne Université, F-75005 Paris, France.*
*[9]FRIPON (Fireball Recovery and InterPlanetary Observation) and Vigie-Ciel Team, France.*
*[10]GEOPS-Géosciences, CNRS, Université Paris-Saclay, 91405, Orsay, France.*
*[11]Service Informatique Pythéas (SIP) CNRS - OSU Institut Pythéas - UMS 3470, Marseille, France.*
*[12]Aix Marseille Univ, CNRS, CNES, LAM, Marseille, France.*
*[13]INAF - Osservatorio Astrofisico di Torino - Via Osservatorio 20, 10025 Pino Torinese, TO, Italy.*
*[14]Astronomical Institute of the Romanian Academy, Bucharest, RO-040557, Romania.*
*[15]MOROI (Meteorites Orbits Reconstruction by Optical Imaging) Astronomical Institute of the Romanian Academy, Bucharest, Romania.*

[16]*SCAMP (System for Capture of Asteroid and Meteorite Paths), FRIPON, UK.*
[17]*Planétarium Rio Tinto Alcan / Espace pour la vie, Montréal, Québec, Canada.*
[18]*Réseau DOME, (Détection et Observation de Météores / Detection and Observation of Meteors).*
[19]*FRIPON-Belgium.*
[20]*Royal Belgian Institute for Space Aeronomy, Brussels, Belgium.*
[21]*Natural History Museum, Burgring 7, A-1010 Vienna, Austria.*
[22]*FRIPON-Austria.*
[23]*Università degli Studi di Torino, Dipartimento di Fisica, Via Pietro Giuria 1, 10125 Torino, TO, Italy.*
[24]*INAF - Osservatorio di Astrofisica e Scienza dello Spazio Via Piero Gobetti 93/3, 40129 Bologna, BO, Italy.*
[25]*INAF - Istituto di Astrofisica e Planetologia Spaziali Via del Fosso del Cavaliere 100, 00133 Roma, RM, Italy.*
[26]*CNR - Istituto di Fisica Applicata Nello Carrara, Via Madonna del Piano, 10 50019 Sesto Fiorentino (FI), Italy.*
[27]*Space sciences, Technologies Astrophysics Research (STAR) Institute, Université de Liège, Liège B-4000, Belgium.*
[28]*Università degli Studi di Firenze - Dipartimento di Scienze della Terra, Via Giorgio La Pira, 4, 50121 Firenze, FI, Italy.*
[29]*Natural History Museum, Cromwell Road, London, UK.*
[30]*Dep. Física Aplicada I, Escuela de Ingeniería de Bilbao, Universidad del País Vasco/Euskal Herriko Unibertsitatea, 48013 Bilbao, Spain.*
[31]*Aula EspaZio Gela, Escuela de Ingeniería de Bilbao, Universidad del País Vasco/Euskal Herriko Unibertsitatea, 48013 Bilbao, Spain.*
[32]*FRIPON-Netherlands, European Space Agency, SCI-SC, Keplerlaan 1, 2201 AZ Noordwijk, Netherlands.*
[33]*Osservatorio Astronomico del Righi, Via Mura delle Chiappe 44R, 16136 Genova, GE, Italy.*

**These authors contributed equally to this work*

Esther.Drolshagen@uni-oldenburg.de; Theresa.Ott@uni-oldenburg.de

## Abstract

In meteor physics the luminous efficiency $\tau$ is used to convert the meteor's magnitude to the corresponding meteoroid's mass. However, lack of sufficiently accurate verification methods or adequate laboratory tests leave this parameter to be controversially discussed. In this work meteor/fireball data obtained by the *Fireball Recovery and InterPlanetary Observation Network (FRIPON)* was used to calculate the masses of the pre-atmospheric meteoroids which could in turn be compared to the meteor brightnesses to assess their luminous efficiencies. For that, deceleration-based formulas for the mass computation were used. We have found $\tau$-values, as well as the shape change coefficients, of 294 fireballs with determined masses in the range of $10^{-6}\ \mathrm{kg} - 100\ \mathrm{kg}$. The derived $\tau$-values have a median of $\tau_{median}$ = 2.17 %. Most of them are on the order of $0.1\ \% - 10\ \%$. We present how our values were obtained, compare them with data reported in the literature, and discuss several methods. A dependence of $\tau$ on the pre-atmospheric velocity of the meteor, $v_e$, is noticeable with a relation of $\tau = 0.0023 \cdot v_e^{2.3}$. The higher luminous efficiency of fast meteors could be explained by the higher energy released. Fast meteoroids produce additional emission lines that radiate more efficiently in specific wavelengths due to the appearance of the so-called second component of higher temperature. Furthermore, a dependence of $\tau$ on the initial meteoroid mass, $M_e$, was found, with negative linear behaviour in log-log space: $\tau = 0.48 \cdot M_e^{-0.47}$. This implies that the radiation of smaller meteoroids is more efficient.

### 1. Introduction

Meteor physics is an old discipline, but, driven by new technology the field has received new interest and current research is producing better and more accurate results now. However, most of the findings in meteor physics still have relatively large uncertainties compared to terrestrial research areas. This is mainly due to approximations of some parameters, such as the shape of the meteoroid and its mass which are unknown for each specific meteor and cannot be measured directly. In addition, both parameters change during the flight through the atmosphere in an unknown, difficult to model, way. The composition of the meteoroid and the atmosphere have to also be approximated, as well as some other aspects of the detection themselves. Additionally, the meteoroid's speed, its height and the meteor's brightness are the most important parameters which are not error-free since they are determined from measurement data with inherent uncertainties. Moreover, the role of the possible fragmentation of the meteoroid along its path should not be disregarded.

Meteors can be observed with various methods e.g. optically, by radar, or by infrasound (see e.g. Brown et al. (2013) or Ott el al. (2020)). The observation equipment determines the meteoroid size range to be recorded. With a small field of view, a high-resolution imaging usually goes hand in hand with many rather faint meteors. Since bright fireballs (visual magnitude $mag$ < -4) are relatively rare, a large field of view is helpful to capture them. However, the larger field of view is usually at the expense of the resolution of the images. A wide FOV together with a favourable spacing between network nodes allows the detection of bright, rare fireballs from more stations than just one. This makes scientific analyses of quite good quality possible. There are different meteor and fireball networks spread around the world, such as the European Fireball Network (Oberst et al. 1998), the Spanish Meteor Network (Trigo-Rodríguez et al. 2004, 2006), the Slovak Video Meteor Network, see e.g. Toth et al. (2012), the Canadian Automated Meteor Observatory, see e.g. Weryk et al. (2013), or the Desert Fireball Network in Australia, see e.g. Howie et al. (2017). The main focus of this paper is on FRIPON, the Fireball Recovery and InterPlanetary Observation Network. It is a French based network that covers France entirely, but also parts of the neighbouring countries, and is currently extending into the rest of Europe. Additional cameras have already been deployed in other countries worldwide. It is designed for fireball detection and meteorite recovery, consisting of all-sky cameras, see Colas et al. (2020).

One goal of many meteor or fireball observations is to compute, together with the trajectory, the pre-atmospheric mass of the corresponding meteoroid. It is usually one of the main parameters to be determined. To do so, often the connection between the entering body's brightness, velocity, and mass is used. This analysis is in turn based on photometric formulae. To compute the pre-atmospheric meteoroid mass from the meteor's brightness, it is assumed what fraction of the loss of kinetic-energy of the object is converted into its brightness. This is described by the parameter $\tau$, the luminous efficiency. The value of $\tau$ is part of various studies as it is a very important parameter in meteor research. Different methods to derive the luminous efficiency have been applied and published in e.g. Verniani (1965), Ceplecha & McCrosky (1976), Halliday et al. (1996), Hill et al. (2005), and Weryk & Brown (2013). These studies all found a dependency of the luminous efficiency on the velocity of the impacting object. Overviews are given e.g. in Koschny et al. (2017) or Subasinghe et al. (2017). Even small variations in the value for $\tau$ can yield very different results for the meteoroid's mass as shown in the two studies just mentioned. This is especially frustrating as the determination of $\tau$ is usually dependent on several assumptions for unknown parameters.

An alternative way to compute the pre-atmospheric mass of the meteoroid is to use the meteor's altitude and its deceleration rate in the atmosphere. This way the mass can be computed regardless of the luminous efficiency. This was done previously e.g. by Gritsevich (2008).

In the just mentioned study, Gritsevich (2008) introduces a dimensionless coefficient method with two parameters: $\alpha$ and $\beta$, which are the ballistic coefficient and the mass loss coefficient, respectively. These values can be computed mainly using the generally more precisely determined velocity and height information for the meteor event. Computations based on this method have already made some meteorite sample recoveries possible as shown in the case of the Annama meteorite, see e.g. Gritsevich et al. (2014a, b). The method was later applied to study, amongst others, the historical *MORB* database to infer the degree of deepening of the fireballs in the atmosphere (Moreno-Ibáñez et al. 2018).

Samson et al. (2019) take advantage of the fact that $\alpha$ and $\beta$ are linked to physical properties of the meteoroid, such as its shape, composition, ablation efficiency, and the duration of the associated meteor. Hence, $\alpha$ and $\beta$ are more than only fitting parameters. For the identification of events with likely meteorite fragments to be found on the ground, Samson et al. (2019) investigated a total of 278 events observed by the Desert Fireball Network allowing them to draw an $\alpha-\beta$ diagram. Using this diagram, they were able to define regions of events for a given minimum terminal mass considering the meteoroid rotation. They classified three distinct regions: events that are unlikely to have dropped a meteorite on the ground, events with possible meteorite falls, and events for which some remnants of the meteoroid are likely to have survived the atmospheric entry and made it to the ground.

Moreover, Moreno-Ibáñez et al. (2020) studied the mathematical equivalence between the logarithm of $2 \cdot \alpha \cdot \beta$ and the PE criterion, which was introduced by Ceplecha & McCrosky (1976) to support the identification of the type of meteor. This criterion is, amongst others, a way to link the fireball's trajectory and the characteristics of the composition of the meteoroid. They found that using the combination of $\alpha$ and $\beta$ offers the possibility for a more general formulation with the use of fewer assumptions.

Furthermore, using the $\alpha$ and $\beta$ method, Gritsevich & Koschny (2011) were able to compute the luminous efficiency with only very few assumptions. In their work, their starting point were the drag and mass loss equations. Considering a change in the meteoroid's velocity and mass during its trajectory and taking the geometrical relation along the path of the meteor into account, they were able to solve the formulas for the meteoroid's dynamical behaviour. The results derived this way are then compared to the drag rate and light curve that were observed for the considered meteor. Afterwards, $\tau$ was computed based on this comparison (Gritsevich & Koschny, 2011).

In a similar approach, Subasinghe et al. (2017) computed the luminous efficiency for simulated meteors using classical meteor ablation equations. The simulations were performed following the model of Campbell-Brown & Koschny (2004). Subasinghe et al. (2017) discussed the uncertainties of $\tau$ by varying the drag coefficient, the meteoroid density, and the shape factor and inferred errors of factors around two for these parameters. In their study, they mention that those parameters were assumed to be constant during the flight through the atmosphere for every simulated meteor. Furthermore, by modelling different masses and velocities of the meteors, they were able to find an uncertainty of a factor of two. The largest deviations were found for the slowest meteors (Subasinghe et al. 2017). In a follow-up publication, Subasinghe & Campbell-Brown (2018) applied the same approach to 15 meteors recorded by the Canadian Automated Meteor Observatory. Characteristics of the network are detailed in e.g. Weryk et al. (2013). The masses used for the simulated meteoroids were on the order of $10^{-6}\,\mathrm{kg} - 10^{-4}\,\mathrm{kg}$ (Subasinghe et al. 2017) and the recorded events had masses in the same size range. They studied small and non-iron meteoroids. They found luminous efficiencies for the non-fragmenting meteoroids between $(0.04 \pm 0.06)\ \%$ and $(27.84 \pm 22.38)\ \%$ with mostly $\tau$ < 1 %. For the fragmenting meteoroids they found upper limits for $\tau$ of a few tens of per cent (see their table 4). Furthermore, they found a weak relationship between the luminous efficiency and the initial meteoroid mass, with negative linear behaviour and no obvious relation between the luminous efficiency and the initial velocity of the meteor. For the relation of luminous efficiency and initial meteoroid mass, they found a linear dependency in log-log space. Converted to the formalism used in this work, their values can be given as ($\tau = b \cdot M_e^a$) $a$ = −0.3647 and $b$ = $0.0016$ (mass $M_e$ in kilograms see their Fig. 10). A comparison with previously published studies placed their $\tau$-values more in the lower value range but they also show large scattering. However, only 12 meteors were taken into consideration in this study, with the limitations that such a small number of meteors analysed imply (Subasinghe & Campbell-Brown, 2018).

Similar results were found by Capek et al. (2019) who investigated double station video observations. In their study they report on observation and modelling of the light curves of meteors produced by small iron meteoroids. Hence, they focussed on faint, slow, and low altitude meteors. They found no obvious relationship between the luminous efficiency and the initial velocity of the meteor, and only a weak relationship between the luminous efficiency and the initial meteoroid mass, with similar negative linear behaviour as in Subasinghe & Campbell-Brown (2018). The values for the luminous efficiency of the 53 studied meteors range from 0.08 % to 5.8 %, and they found the relation $\tau = \beta \cdot M_e^{\alpha}$ with $\beta = 2.0^{+1.0}_{-0.7}$, $\alpha = -0.38 \pm 0.11$, with the initial meteoroid mass, $M_e$, in milligrams (Capek et al. 2019).

It is also obvious that to infer the luminous efficiency of small meteors poses an intrinsic challenge since they exhibit compositional differences (Trigo-Rodríguez et al. 2003). As stream meteoroids are produced by the decay of undifferentiated asteroids and comets, they are composed of fine-grained aggregates (probably similar to Interplanetary Dust Particles, IDPs) that are built by diverse minerals. The random distribution of such mixtures might produce different bulk chemical compositions and tensile strengths (Rietmeijer 2004; Trigo-Rodríguez & Blum 2009). As a consequence, the meteor ablation columns could have varying chemical elements to be excited during the meteor ablation, so it should produce significant luminous efficiency differences (Borovička & Spurný 1996; Trigo-Rodríguez et al. 2003). Moderately volatile elements are depleted at higher altitudes where excitation potentials are lower, producing the so-called differential ablation (Trigo-Rodríguez et al. 2004; Gómez-Martín et al. 2017). At the same time, faster meteoroids develop a second spectral component that increases the luminosity of their meteors (Borovička 1994; Trigo-Rodríguez et al. 2003). To better understand the changes in luminous efficiency it is important to remember that meteor radiation is not continuous but produced by emission lines that lie in different regions of the electromagnetic spectrum. Just as an example, fast meteors containing silicates will show the ionized Si II lines at 408 nm and 635 nm, this way increasing the luminous efficiency (Trigo-Rodríguez 2019). Silicon is a very common chemical element as it forms part of the omnipresent silicates in chondritic materials (Trigo-Rodríguez et al. 2019).

This work focusses on the luminous efficiency computation using the deceleration-based method following the procedure of Gritsevich & Koschny (2011), described in detail below, in Section 2. For the fireballs observed by FRIPON, data are presented in Section 3, whereas the computation of luminous efficiencies is presented in Section 5. The results are shown in Section 6. In Section 4 other published studies are briefly presented. Finally, in Section 7 our results are compared to those from other studies. Additionally, in Section 7 relations between the luminous efficiencies and the objects' initial masses and velocities as well as entry angles are investigated and presented, followed by a conclusion.

## 2. Deceleration-based computations

To compute the pre-atmospheric mass of a meteoroid, a photometric relationship is used in most common methods. It is assumed that during the deceleration in the Earth's atmosphere a certain fraction of the loss of kinetic energy of the meteoroid is converted into its brightness. How large this fraction is, is part of extensive studies with varying values for this luminous efficiency. Using these differing values published in literature can lead to results with a large variation in the mass of the corresponding meteoroid, as shown in e.g. Koschny et al. (2017).

For each recorded fireball event, we have some event specific parameters, such as, but not limited to, the pre-atmospheric mass and velocity, i.e. $M_e$ and $v_e$, the entry angle $\gamma$ but also some calculated values such as the luminous efficiency $\tau$. Some parameters, however, are not only event specific but also time dependent, measured over the course of the trajectory. These parameters are e.g. the intensity $I$ and the velocity $v$ of the meteor. Furthermore, some parameters will be introduced which were divided by the corresponding pre-atmospheric value and are dimensionless, e.g. the dimensionless velocity $v^*$ and mass $m^*$. These event parameters are marked with an asterisk.

In an effort to remove the effect of the luminous efficiency on the mass computation Gritsevich (2008), amongst other authors, developed an approach to compute the pre-atmospheric mass of the meteoroid by using the altitude of the meteor and its rate of deceleration in the atmosphere. This way a luminous efficiency does not have to be assumed for its computation. This method is briefly explained here.

Gritsevich (2008) computes the pre-atmospheric mass of the meteoroid by using the meteor's altitude and the rate of deceleration. This is done by computing the ballistic coefficient $\alpha$ describing the deceleration rate and using the mass loss parameter $\beta$ (compare Gritsevich (2008) eq. (8) and (9)):

$$\alpha = \frac{c_d \cdot \rho_0 \cdot h_0 \cdot S_e}{2 \cdot M_e \cdot \sin(\gamma)} \quad (1)$$

$$\beta = \frac{(1-\mu) \cdot c_h \cdot v_e}{2 \cdot c_d \cdot H} \quad (2)$$

with the drag coefficient $c_d$, the gas density at sea level $\rho_0$, the pre-atmospheric cross section area of the meteoroid $S_e$, the scale height $h_0$, the pre-atmospheric meteoroid mass $M_e$, the angle between horizon and trajectory $\gamma$, the heat-transfer coefficient $c_h$, the pre-atmospheric meteoroid velocity $v_e$, the effective destruction enthalpy $H$, and the shape change coefficient $\mu$. The latter represents the effect of the change of the object's shape, ranging from 0 to 2/3. A value of $\mu = 0$ corresponds to the case where the maximal heating and evaporation occurs in the front of the meteoroid, $\mu = 2/3$ represents a uniform mass loss of the meteoroid over the entire surface. Most studies speculate that $\mu$ is related to the meteoroid's rotation (e.g. Gritsevich & Koschny (2011) or Samson et al. (2019)).

Based on the drag and mass loss equations, Gritsevich (2008) derived a formula dependent on the dimensionless parameters $\alpha$ and $\beta$, as well as on the dimensionless velocity $v^*$, see Eq. (4) (Gritsevich (2008) eq. (7)). This equation represents a height-velocity relation for which $\alpha$ and $\beta$ represent the best solution. Since $v^*$ depends only on the deceleration of the meteoroid, it is possible to determine the proper values of $\alpha$ and $\beta$ obtained from the best least-square fit of the observed heights and velocities for equation (4).

This method has the main advantage that no assumptions of the meteoroids' parameters have to be made. The parameters $\alpha$ and $\beta$ can be derived by using only the generally accurately measurable altitude of the meteoroid and its velocity.

$$m^* = \exp\left(-\beta \cdot \frac{1-v^{*2}}{1-\mu}\right) \quad (3)$$

$$y^* = \ln(\alpha) + \beta - \ln\left(\frac{\overline{Ei}(\beta) - \overline{Ei}(\beta \cdot v^{*2})}{2}\right), \quad (4)$$

with the exponential integral $\overline{Ei}(x)$:

$$\overline{Ei}(x) = \int_{\infty}^{x} \frac{e^z dz}{z}, \quad (5)$$

and the dimensionless velocity, mass, and height:

$$v^* = \frac{v}{v_e}, \qquad m^* = \frac{M}{M_e}, \qquad y^* = \frac{h}{h_0} \quad (6)$$

Using Eq. (7) (Gritsevich (2008) eq. (12)), the pre-atmospheric meteoroid mass can be derived. It originated from Eq. (1) converted to $M_e$ with the pre-atmospheric shape factor of the meteoroid $A_e$ and the meteoroid bulk density $\rho$. In addition, to estimates for $A_e$ and $c_d$ an assumption for $\rho$ is needed. The scale height of the Earth's atmosphere $h_0$ is expected to be 7160 m and the gas density at sea level $\rho_0$ is 1.29 kg/m$^3$.

$$M_e = \left(\frac{1}{2} c_d \frac{(\rho_0 h_0)}{\alpha \sin(\gamma)} \frac{A_e}{\rho^{2/3}}\right)^3 \quad (7)$$

In Gritsevich & Koschny (2011), they use a computation procedure for a fireball's brightness and luminous efficiency depending on the mass and velocity of its source meteoroid building on the work by Gritsevich (2008). They derived Eq. (8) and Eq. (9) (their eq. (13) and (14)). In these equations it can be seen that since $\beta$ and $M_e$ could be derived as explained in Gritsevich (2008), $\mu$ and $\tau$ remain the unknown variables in the equations. The proper value of $\mu$ and $\tau$ can then be found by comparing the shape of the observed light curve and applying a least-square fit with equation (8).

$$I(v^*) = \frac{\tau \cdot M_e \cdot v_e^3 \cdot \sin(\gamma) f(v^*)}{2 \cdot h_0} \quad (8)$$

with

$$f(v^*) = v^{*3} \cdot \left(\overline{Ei}(\beta) - \overline{Ei}(\beta \cdot v^{*2})\right) \cdot \left(\frac{\beta \cdot v^{*2}}{1-\mu} + 1\right) \cdot \exp\left(\frac{\beta \cdot (\mu \cdot v^{*2} - 1)}{1-\mu}\right), \quad (9)$$

with the shape change coefficient $\mu$ and the luminous efficiency $\tau$. The sole remaining unknown parameter in this equation is the meteor brightness $I$, which is based on observations. Hence, $\mu$ and $\tau$ are derived by finding the best fit of the brightness function to photometrical observations (Gritsevich & Koschny, 2011).

To summarize, the specific steps to derive the luminous efficiency of a meteoroid from data of an observed fireball are the following:

1. Obtain the heights and velocities of the fireball
2. Obtain the optimal values of $\alpha$ and $\beta$ from the least-square fit of the observed heights and velocities according to Eq. (4)
3. Determine the mass $M_e$ according to Eq. (7) with assumptions for $A_e$, $c_d$, and $\rho$
4. Obtain the brightness values of the fireball
5. Obtain the optimal values of $\mu$ and $\tau$ from the least-square fit of the observed light curves and velocities according to Eq. (8)

In the course of the present work, only steps 3 to 5 were carried out. In Section 4 a specific example of the method is presented.

## 3. FRIPON data

The Fireball Recovery and InterPlanetary Observation Network (FRIPON) is a global fireball network which started in France in 2016. These days it not only covers France, but also includes several of the neighbouring European countries and is expanding worldwide. Consisting of all-sky cameras, it is optimized for fireball detection and meteorite recovery. The stations have an average distance from each other of around 80 km, to allow triangulation measurements of the recorded fireballs. A fireball that is observed by at least two stations is detected automatically and stored in the FRIPON database. As of 2020 May, the network consists of 150 cameras and 20 radio antennas, supplementing the optical observations. This is a total covered area of about $1.5 \cdot 10^6$ km$^2$ (Colas et al. 2020). For more details about the network, its distribution, and the collaborating countries we refer to Colas et al. (2020). The software used for the control of the camera and meteor event detection is called FreeTure, see Audureau et al. (2014) for details.

The stations design is quite simple, which is one of the advantages of the network, opening it up to institutions and amateurs alike. The digital cameras that are used are DMK 23G445 cameras (Anghel et al. 2019) based on the Sony chip ICX445 with 1.2 megapixels and a pixel size of 3.75 μm x 3.75 μm (Colas et al. 2014, 2020). For now, they operate only during night time with 30 frames per second and take one 5 seconds long exposure every 10 minutes for calibration purposes. The fish-eye lens has a $f = 1{:}25$ mm (Jeanne et al. 2019). Furthermore, FRIPON uses a hardware configuration of a NUC with i3 processor, 8 Gb of RAM (for image buffering), 32 Gb SSD for system installation, 1 Tb HDD, GigE Vision, and PoE (Power over Ethernet) (Colas et al. 2014, 2020). Since the system has been further developed during the past few years, more recent stations can have a slightly different hardware configuration. For further information about FRIPON, see Colas et al. (2020).

Most countries have integrated their national network into the FRIPON database, e.g. the Italian network, called PRISMA (Gardiol et al. 2016, 2019), SCAMP in the UK, as well as e.g. the networks of Germany, the Netherlands, or Spain. Some national networks are still in the installation phase and will also be integrated into the FRIPON database in the future, like the MOROI network covering Romania (Nedelcu et al. 2018; Anghel et al. 2019).

Recently, first statistics and results of the FRIPON data were published in Colas et al. (2020). Since 2016, around 3700 fireballs were detected by the network. Taking the increasing number of installed cameras into account, the network observes on average about 1000 meteors per year of which 55 % were classified as sporadic meteors. The accuracy of the astrometric reduction is estimated to be 1 arcmin and the velocity values have uncertainties of about 100 m/s. Furthermore, the accuracy of the photometry is around 0.5 mag. This value is only valid for non-saturated events with an absolute magnitude of at most -8. Additionally, FRIPON is not fully efficient for events with an absolute magnitude fainter than -5 mag which corresponds to a detection threshold for the incoming meteoroids of objects with a size of about 1 cm (Colas et al. 2020).

## 4. Luminous Efficiencies in the Literature

A frequently used method to compute the pre-atmospheric mass of a meteoroid, as previously mentioned, is based on a photometry formula. The conversion between the brightness of a recorded meteor or fireball to the mass of the corresponding meteoroid is a complex topic involving multiple dependencies and parameters. Many authors have already addressed this problem in various publications, using different assumptions. In most cases, the brightness of the meteor is integrated along its visible trajectory. The amount of kinetic energy released by deceleration of the entering object in our atmosphere is converted into visible radiation and emitted as light intensity, $I$, must be estimated. This fraction is called the luminous efficiency $\tau$ and used to determine the pre-atmospheric meteoroid mass from the measured brightness of the meteor.

Following Verniani (1965) the relation can be given as:

$$I = -\frac{\tau \cdot v_e^2}{2}\frac{dM}{dt}, \quad (11)$$

with the meteoroid's mass loss $dM/dt$. Hence, the pre-atmospheric meteoroid mass $M_e$ can be computed as

$$M_e = \frac{2}{\tau \cdot v_e^2}\int I_s \, ds, \quad (12)$$

with the term $\int I_s \, ds$ giving the emitted light $I$ in Watts, integrated over the complete flight path $s$.

$I$ can be expressed as shown in Eq. (13), wherein $P_{0M}$ is the radiant power of a zero-magnitude meteor and dependent on the observation set up. $Mag$ is the absolute magnitude of the meteor at peak brightness (Drolshagen & Kretschmer, 2015). The equation has to be modified for the observational setup.

$$I = P_{0M} \cdot 10^{-0.4 \cdot Mag} \quad (13)$$

The value for $\tau$ is difficult to determine since it depends on parameters which usually are unknown for the specific meteor and cannot be measured directly. Hence, they have to be estimated with quite some uncertainties. Two of these parameters are the shape of the meteoroid and its mass. The change and evolution of these parameters during the flight of the object through the atmosphere are usually also unknown. In addition, a possible fragmentation of the meteoroid should be taken into account as well. Moreover, the composition of the meteoroid and the atmosphere have to be approximated. Additionally, some aspects of the detection themselves have to be taken into consideration. Other significant parameters like the meteoroid's velocity, its height and the meteor's brightness are also not error-free since they are based on measurements with inherent uncertainties.

Several authors studied the luminous efficiency and published a relation for the luminous efficiency and the meteor's velocity. Since different assumptions were required depending on the method used in these studies, a direct comparison is difficult. Several factors have to be considered. Most studies do not take fragmenting meteoroids into account, even though it is known that fragmentation can have a large effect as shown by e.g. Subasinghe & Campbell-Brown (2018) or Ceplecha & ReVelle (2005). In fact, most meteoroids fragment as shown by e.g. Subasinghe et al. (2016). The recorded light curve is highly dependent on the detecting instrument used and its interpretation of the assumed blackbody source function. Furthermore, the spectral sensitivity of the instrument affects the received photon radiant power. As the spectra of meteors are not homogeneous (especially since there are various discrete emission lines present) the wavelength sensitivity of the recording instrument influences the detection capability and probability. Borovicka et al. (1999) showed a continuum for video meteors as well as diverse differences in emissions of the analysed meteors. Additionally, the assumed composition of the object that caused the observed meteor has a large effect on $\tau$, as shown by e.g. Svetsov & Shuvalov (2018).

Verniani (1965) published a formula for the luminous efficiency investigating meteor data from the Harvard photographic meteor project. He studied data of 413 Super-Schmidt meteors with magnitudes in the range between about – 1 and + 0.5. The initial masses were mentioned to be on the order of $10^{-2}$ kg. Verniani (1965) found a linear relationship between $\tau$ and the velocity, with values for $\tau$ on the order of about $0.02\ \% - 0.2\ \%$, after taking into account that the calculated values are in zero magnitude and cgs units.

Photographic fireball data of 232 observations collected by the Prairie networks were investigated by Ceplecha & McCrosky (1976). Moreover, they combined their results with the ones by Friichtenicht et al. (1968) who carried out laboratory measurements of $\tau$ and with the ones by Ayers et al. (1970), who carried out an analysis of artificial meteors. Ceplecha & McCrosky (1976) computed masses of their analysed meteors in the mass range of $10^{-1}$ kg - $10^{2}$ kg and a velocity dependent relation of $\tau$. Taking into account that the values are returned in zero magnitude and cgs units the range of the luminous efficiency was of ca. $0.2\ \% - 2\ \%$.

From an analysis of 259 fireballs detected by a Canadian camera network a velocity dependent relation for $\tau$ was derived by Halliday et al. (1996). They analysed objects in the mass range of $10^{-3}$ kg - $10^{2}$ kg and found $\tau$-values in the range of about 1 % - 7 %.

The focus of the numerical work by Hill et al. (2005) was on small ($10^{-13}$ kg - $10^{-6}$ kg) high-velocity meteors. Based on the theoretical investigation by Jones & Halliday (2001) they published a formula for $\tau$ with a velocity dependent extinction coefficient. Taking the spectral band as well as the chondritic ratio into account (as described in Campbell-Brown et al. (2012)) they give a range of the luminous efficiency of about $0.2\ \% - 0.7\ \%$.

Weryk & Brown (2013) combined data from simultaneous detections by the Canadian meteor radar system and optical video camera systems. By combining the observations, they were able to find a relation between the ratio of the ionisation coefficient to the luminous efficiency $\tau$ and the meteoroid velocity. The analysed objects were in the mass range of about $10^{-8}$ kg - $10^{-5}$ kg and they found luminous efficiencies in the range of about $0.5\ \% - 6\ \%$.

In Gritsevich & Koschny (2011) the theoretical concept of computing $\tau$ described in Section 2 and used in this work was applied to three fireballs of the dataset published by Halliday et al. (1996). The objects' pre-atmospheric masses were 5.6 kg, 2.4 kg, and 87 kg. They found luminous efficiencies in the range of about $0.6\ \% - 8\ \%$.

Nemtchinov et al. (1997) analysed data of 51 light flashes detected from space by satellites' optical sensors. With numerical simulations they computed theoretical values of luminous efficiencies for objects in the size range of 0.2 m – 20 m with, amongst others, different densities. They found an increase of the luminous efficiency with growing size as well as with increasing initial velocity of the entering body. They published a mean value of $\tau = 5\,\% - 10\,\%$.

The spectrum of fireballs depends on various factors like its composition, size, velocity, or height. The sensitivity of different types of detectors varies as well for different wavelengths. This is an important factor to be considered when comparing results derived from different systems' data. Especially when evaluating results derived from data recorded by ground based optical systems and by space based systems.

Svetsov & Shuvalov (2019) used numerical simulations to investigate atmospheric impacts of stony and icy cometary objects in the size range of 0.3 km to 3 km, taking their velocity and entry angle into account. They found values for $\tau$ from 1 % - 18 % (bolometric values). Most of the investigated objects were quite large and not significantly decelerated by the atmosphere. This is especially important since they found an increase in the luminous efficiency for smaller sizes likely due to deceleration close to the ground in a previous work (Svetsov & Shuvalov 2018). For a size range between 30 m and 100 m they found bolometric luminous efficiencies of some percent up to 20 % for asteroids and 40 % for icy bodies. Svetsov & Shuvalov (2018) concluded that the maximum mean luminous efficiency should be reached by objects with a size between 10 m and 30 m. For smaller objects they expect a linear relation of the luminous efficiency and mass. Furthermore, they found a weak dependency of luminous efficiency on pre-atmospheric velocity as well as a relatively strong relation with the entry angle. Especially objects with very shallow angles show large luminous efficiencies in their simulations. Svetsov & Shuvalov (2018) mention that the luminous efficiency in the visible range lies between 30 % and 50 % of the total luminous efficiency, which has to be taken into consideration if one compares results from different sensors. Following their fig. 4, the presented luminous efficiency for the range of 1.6 eV – 3.2 eV (about 390 nm - 775 nm) is between 1 % and 19 %.

It should be mentioned, that Nemtchinov et al. (1997), Svetsov & Shuvalov (2018), and Svetsov & Shuvalov (2019) define the luminous efficiency as the fraction of the kinetic energy of the impacting object that is radiated. Nemtchinov et al. (1997) mentioned that this might not correspond entirely to the luminous efficiency as defined by the fraction of the kinetic energy of the impacting object that is transformed into light and radiated during the object's passage through the Earth's atmosphere due to ablation and deceleration of the meteoroid. Which is a fraction of the object's total energy lost. Svetsov & Shuvalov (2018) mention that it is unknown which kinetic energy losses are linked to the radiation flux for large objects with sizes on the order of tens of metres. They discuss that for some time the kinetic energy of the impacting objects is retained by the vapour generated by the energy transfer. But, after a complete vaporization of the object its radiation of light is still intense, for low heights in combination with low velocities. At this stage, its kinetic energy could have decreased to about 1 % of its initial kinetic energy. This would cause a calculated luminous efficiency tending towards infinity.

Ceplecha & ReVellle (2005) developed a numerical fragmentation model and applied it to observed fireball events. They distinguished between the apparent and the intrinsic luminous efficiency. They defined the apparent values as the ones that are not corrected for fragmentation, hence for which a single body was assumed. For the intrinsic luminous efficiency they take effects of fragmentation into account by using their fragmentation model. They found substantial changes of the determined apparent luminous efficiencies during the trajectory of the investigated events. Additionally, the determined intrinsic luminous efficiencies are much smaller than the corresponding apparent values. Furthermore, for some of these events, Ceplecha & ReVellle (2005) found values of the apparent luminous efficiencies on the order of 100 %, at least for parts of the trajectories, and an increasing difference between the two $\tau$-values with increasing end height. They concluded that the primary process of the ablation of the entering objects is fragmentation. Since the apparent luminous efficiencies neglect the effects of fragmentation, they classify them as rather unrealistic. By not considering fragmentation, they also explain the difference between the photometric mass and the dynamical mass.

The luminous efficiency of the Chelyabinsk asteroid was estimated based on observations by Brown et al. (2013) to be about 17 % for the whole wavelength range. The object had a diameter of 19 m and an entry angle of approximately 19°. Svettsov et al. (2018) carried out numerical simulations of this event publishing 17 % as the luminous efficiency of the entire spectrum, the ratio of total radiated energy to the kinetic energy. The luminous efficiency in the wavelength range between 350 nm and 650 nm was found to be of about 6 %. Again, this has to be taken into account when comparing e.g. photographical observations and data collected by satellites.

For the Innisfree meteorite fall from 1977 Halliday et al. (1981) computed values for the luminous efficiency in the range of 4 % - 8 %, depending on the assumed entry mass of the object of 20 kg or 40 kg. Their study was based on the recovered meteorite fragments and photographic observations with dynamical data.

As already described in the introduction Subasinghe & Campbell-Brown (2018) studied the luminous efficiency of small, non-iron meteoroids with masses on the order of $10^{-6}\,\mathrm{kg} - 10^{-4}\,\mathrm{kg}$. They found luminous efficiencies for the non-fragmenting meteoroids ranging from about 0.04 % up to 30 %, whereas most of those had luminous efficiencies less than 1 %. For the fragmenting meteoroids they found upper limits for $\tau$ of a few tens of percent. Furthermore, they found a weak relationship between the luminous efficiency and the initial meteoroid mass with negative linear behaviour.

Capek et al. (2019) investigated double station video observations of faint, slow, and low altitude meteors produced by small iron meteoroids of masses in the milligrams range. They found a relationship between the luminous efficiency and the initial meteoroid mass with similar negative linear behaviour as Subasinghe & Campbell-Brown (2018). They published values for the luminous efficiencies in the range of about 0.08 % – 6 %, and found the relation $\tau = \beta \cdot M_e^{\alpha}$ with $\beta = 2.0^{+1.0}_{-0.7}$, $\alpha = 0.38 \pm 0.11$ (Capek et al. 2019).

An overview of the mentioned studies is given in Table 1.

*Table 1: Overview of luminous efficiency studies with mass range of the investigated objects, utilized data sources and derived range of $\tau$-values. For those works which have indicated their size ranges in metres, we have estimated the masses in kilograms, assuming a density of 2500 $kg/m^3$. They are stated in brackets.*

| **literature** | **mass range** | **τ range** | **sources** |
|---|---|---|---|
| Verniani (1965) | order of $10^{-2}$ kg | 0.02 % - 0.2 % | Harvard photographic meteor project, 413 Super-Schmidt meteors, -1 to + 0.5 mag |
| Ceplecha & McCrosky (1976) | $10^{-1}$ kg - $10^{2}$ kg | 0.2 % - 2 % | Photographic fireball data, 232 observations, the Prairie networks + laboratory measurements of τ + |
| | | | artificial meteors (Friichtenicht et al. (1968) + Ayers et al. (1970) were taken into account) |
| Halliday et al. (1996) | $10^{-3}$ kg - $10^{2}$ kg | 1 % - 7 % | 259 fireballs detected by a Canadian camera network |
| Hill et al. (2005 ) | $10^{-13}$ kg - $10^{-6}$ kg | 0.2 % - 0.7 % | numerical, high-velocity meteors |
| Weryk & Brown (2013) | $10^{-8}$ kg - $10^{-5}$ kg | 0.5 % - 6 % | Canadian meteor radar system and optical video camera systems |
| Gritsevich & Koschny (2011) | 1 kg – $10^{2}$ kg | 0.6 % - 8 %. | three fireballs of the sample published by Halliday et al. (1996) |
| Nemtchinov et al. (1997) | 0.2 m – 20 m (10 kg - $10^{7}$ kg) | 5 % - 10 % (bolometric) | light flashes detected from space by satellites' optical sensors |
| Svetsov & Shuvalov (2019) | 0.3 km to 3 km ($10^{10}$ kg - $10^{13}$ kg) | 1 % - 18 % (bolometric) | simulations |
| Svetsov & Shuvalov (2018) | 30 m and 100 m ($10^{7}$ kg - $10^{9}$ kg) | 2 % - 40 % (bolometric) 1 % - 19 % (390 nm - 775 nm) | simulations |
| Brown et al. (2013) | 19 m ($10^{6}$ kg) | 17 % (bolometric) | Chelyabinsk meteoroid |
| Svettsov et al. (2018) | 19 m ($10^{6}$ kg) | 17 % (bolometric) 6 % (350 nm - 650 nm) | Chelyabinsk meteoroid |
| Halliday et al. (1981) | 20 kg - 40 kg | 4 % - 8 % | Innisfree meteorite fall from 1977 |
| Subasinghe & Campbell-Brown (2018) | $10^{-6}$ kg – $10^{-4}$ kg | 0.04 % - 1 % (some up to 30 %) | small, non-fragmenting, and non-iron meteoroids |
| Capek et al. (2019) | $10^{-6}$ kg - $10^{-3}$ kg | 0.08 % – 6 % | double station video observations of faint, slow, and low altitude meteors |
| This work | $10^{-6}$ kg – $10^{2}$ kg | Most: 0.1 % - 10 % (4 % of dataset: $10^{-4}$ % - 0.1 % 26 % of dataset: 10 % - 100 %) | 294 FRIPON fireballs |

## 5. Luminous Efficiency Computation

To illustrate the principle of the computations carried out within the framework of this study, each step is presented in detail below using an example fireball event. Table 2 includes some information of the event that occurred on 2018 May 19 around 01:40 UT above the Occitanie region, France. It was detected with seven different stations. It was chosen as an example, because the single station data of the brightness are well suited to visualize the concept. Furthermore, the (calculated) values and properties are close to the average values of the subset analysed herein and therefore a nice representation of a standard fireball event. An image of the fireball recorded with the FRIPON station located at Onet-le-Château is shown in Fig 1. The pre-atmospheric velocity of the meteoroid, as computed by the FRIPON pipeline, was 17.2 km $s^{-1}$. Furthermore, in Table 2 the computed maximum absolute magnitude of the fireball is shown; for a more detailed explanation please refer to the end of this section. In the last two columns of the Table 2 the shape change coefficient $\mu$ and the luminous efficiency $\tau$, computed following the method of Gritsevich & Koschny (2011) and described in Section 2, are also reported.

The FRIPON pipeline applies a slightly modified version of the method from Gritsevich (2008) to compute the pre-atmospheric masses of the meteoroids. The data processing is described in more detail in Jeanne et al. (2019), essentially, an optimization of a fit through their velocity data is used. It should be mentioned that it is assumed that $\mu = 2/3$ from the very beginning. Even though this value of $\mu = 2/3$ is used by most authors in the literature (i.e. characteristic for meteoroids with surfaces proportional to their volume to the power of 2/3 and heat redistribution due to rotation to the whole surface) it is an uncertainty that should not be overlooked. Nonetheless, $\alpha$ and $\beta$ are only based on deceleration data.

Since we use the method of Gritsevich & Koschny (2011) to process the FRIPON data, we are able to apply our own least-square fit to the observed light curves and find the optimal values of $\mu$ and $\tau$. Depending on the assumed values for $A_e$, $c_d$, and $\rho$, and the obtained value of $\mu$, $M_e$ might differ from the values determined by the FRIPON pipeline.

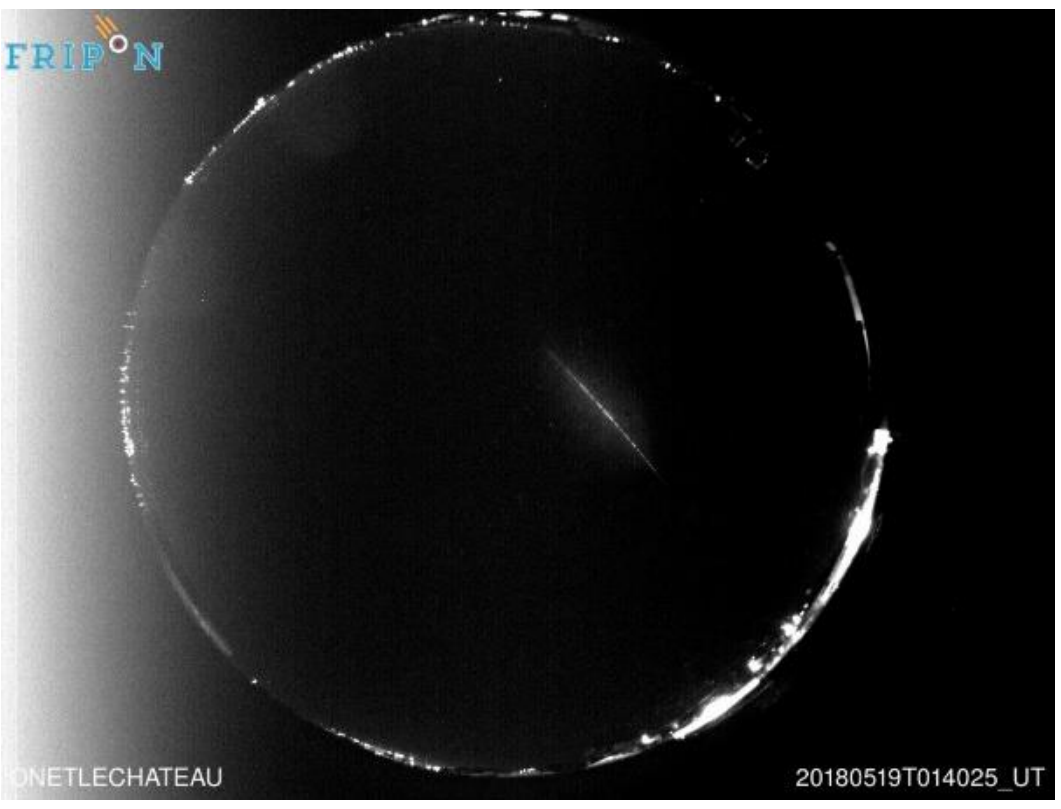

*Figure 1: Fireball from 2018 May 19 as recorded with the FRIPON station of Onet-le-Château (Occitanie region, France).*

For every fireball analysed by FRIPON, the following necessary values were extracted: the mass loss parameter $\beta$, the pre-atmospheric velocity $v_e$, and the slope between horizon and trajectory $\gamma$. The other fireball specific parameter from Gritsevich (2008), the ballistic coefficient $\alpha$, is also computed and listed but is not required for the computation of the luminous efficiency with the applied method.

As it can be seen in Eq. (8) the pre-atmospheric meteoroid mass $M_e$ is needed to compute $\tau$. $M_e$ is determined with Eq. (7). In this step some assumptions have to be made. In literature, these assumptions vary from one study to another.

The equation includes the drag coefficient $c_d$, the pre-atmospheric shape factor of the meteoroid $A_e$, and the meteoroid bulk density $\rho$. As in Gritsevich & Koschny (2011), we assumed $c_d = 1.2$ and $A_e = 1.5$. Another common assumption is an initial spherical shape of the meteoroid, with $A_e = 1.21$ and $c_d = 1$. A review of generally used assumptions and parameters was published by e.g. Gritsevich (2008b). Assumed values for the meteoroid bulk density $\rho$ can span a wide range of values, as discussed in e.g. Gritsevich & Koschny (2011). For about 45 % of the whole FRIPON dataset Colas et al. (2020) found a shower affiliation. Hence, we expect that a lot of fireballs originate from cometary material and that e.g. meteorite densities could at the most be taken as an upper boundary density for a set of recorded meteoroids. The most fragile parts do not survive their way through the atmosphere. The European Cooperation for Space Standardization (ECSS) supposes a meteoroid density of $\rho = 2500$ kg/m$^3$ for impact risk assessments for satellites (ECSS, 2008). Based on this recommendation, we use the ECSS meteoroid density estimation for the computations in this study. Nonetheless, all these parameters do affect the mass and, hence, the results for the luminous efficiency, too. Thus, the varying assumptions for these parameters from one study to another contribute to a large range of luminous efficiency values.

Furthermore, the values computed by FRIPON, e.g. $\alpha$ and $\beta$, are not free of errors. For example, to compute the pre-atmospheric meteoroid mass, see Eq. (7), $\alpha$ is needed. The pre-atmospheric mass is proportional to $\alpha$ by $M_e \sim \alpha^{-3}$. Furthermore, the brightness is proportional to the pre-atmospheric mass, as can be seen in Eq. (8) and hence to $\alpha$ by $I \sim M_e \sim \alpha^{-3}$. If one converts the formula to the luminous efficiency it is also easy to see that $\tau \sim \alpha^3$. Hence, an error in $\alpha$ of a factor of 2 would cause an error in the computations by a factor of 8 in $M_e$ and $\tau$. This emphasizes how large the luminous efficiency is dependent on $\alpha$.

The velocity $v(t)$ of the fireball is given for each calculated point in time of the trajectory. The computation of the trajectory is carried out by the FRIPON pipeline and explained in detail in Jeanne et al. (2019). The meteor's brightness at these points in time is not computed by the FRIPON pipeline but available for the individual station's recorded data of the fireball. For FRIPON's trajectory computations the stations' data are weighted depending on the 'quality' of the single station data, as explained in Jeanne et al. (2019). So far, this is only done for the astrometry and not for the photometry.

In order to calculate the luminous efficiency, we matched the fireball's brightness with the velocity for the same points in time. We also extracted the fireball's absolute magnitude values of each station. This way we receive the absolute magnitude values for all stations for each point in time when the meteor was visible. From these single stations' magnitude data, we exclude outliers and if more than 50 % of a light curve of any station was excluded, the entire light curve is not considered in further calculations. This method is presented in Fig. 2. In the Fig. 2 the interpolated absolute magnitude single station data of our example fireball is presented as color-coded lines. Shown as blue dots are the median absolute magnitude values, calculated for each point in time when the fireball was visible. This curves' data points are obtained as follows: First, from all stations' absolute magnitude values, the outliers at each point in time are excluded by applying the statistical three-sigma rule: all values that do not lie within three standard deviations of the mean value are disregarded (see e.g. Everitt & Skrondal, 2002). All single station data light curves of which more than 50 % of the data were rejected, were also excluded from further calculations. For all remaining data points the three-sigma rule is applied again. This way, parts of the curves with large deviation are discarded as well. In Fig. 2 it can be clearly seen that in this way the data of station FRMP07 as well as parts of FRMP04 will be classified as outliers. If only data of one station is available for parts of the observed time frame this part of the light curve is also rejected. The peak brightness of the resulting median magnitude curve is highlighted in the Fig. 2 with an orange 'x' and listed in Table 2. The error estimations of the luminous efficiency and the shape change coefficient are based on their respective fit to the light curve. Please refer to Jeanne et al. (2019) and Colas et al. (2020) for a discussion on the FRIPON generated parameters' errors.

*Table 2: Details of the 2018 May 19 fireball as detected by FRIPON, including the number of detecting stations, the computed meteor's pre-atmospheric velocity $v_e$, pre-atmospheric meteoroid mass $M_e$, the shape change coefficient $\alpha$, the mass loss parameter $\beta$, the slope between horizon and the trajectory $\gamma$, as well as the determined maximum absolute magnitude, the shape change coefficient $\mu$, and the luminous efficiency $\tau$ derived in this work as explained in Section 2 with Eq. (8) and (9).*

| Date | Time in UT | # Stations | $v_e$ in km s$^{-1}$ | $M_e$ in kg | $\alpha$ | $\beta$ | $\gamma$ in ° | Peak brightness | $\mu$ | $\tau$ in % |
|---|---|---|---|---|---|---|---|---|---|---|
| 2018-05-19 | 01:40:24 | 7 | 17.17 ± 0.03 | 0.52 | 77.26 | 1.21 | 46.7 | -6.7 ± | 0.681 ± 0.003 | 1.495± 0.007 |

Since the luminous efficiency calculation requires the flux $I$ we use our derived magnitude values $Mag$ and compute the brightness with Eq. (10) (compare Eq. (18) in Gritsevich & Koschny (2011), which is based on an equation developed by Ceplecha & ReVelle (2005)):

$$I = 10^{-0.4 \cdot Mag + 3.185} \quad (10)$$

However, some challenges of this method should be considered: In the case of particularly bright events, the cameras will saturate at around -8 mag. However, only the data of individual stations are affected, particularly those located directly below the fireball. Other stations, further away ought to have recorded unsaturated data. The erroneous curves should be recognized as outliers during the calculation of the median curve and then ignored for the rest of the calculation. The dataset used in this study consists of 24 events with peak brightnesses brighter than -7 mag and for which we would consider that they might be affected by saturation effects. Also to be considered are clouds that could occur and would affect the data of individual cameras. A weighting of the single station data based on its quality could enhance the accuracy of the magnitude values. However, this could also introduce sources of error due to necessary assumptions. An error of 1 magnitude would lead to a factor of about 2.5 difference in brightness and hence to a luminous efficiency with the same error, compare Eq. (8). Since the FRIPON cameras are all-sky cameras, the photometric values are expected to have relatively large uncertainties on the order of half a magnitude. Keeping this in mind, we expect the brightness values computed as explained here to be a quite good compromise between accuracy and computational efforts.

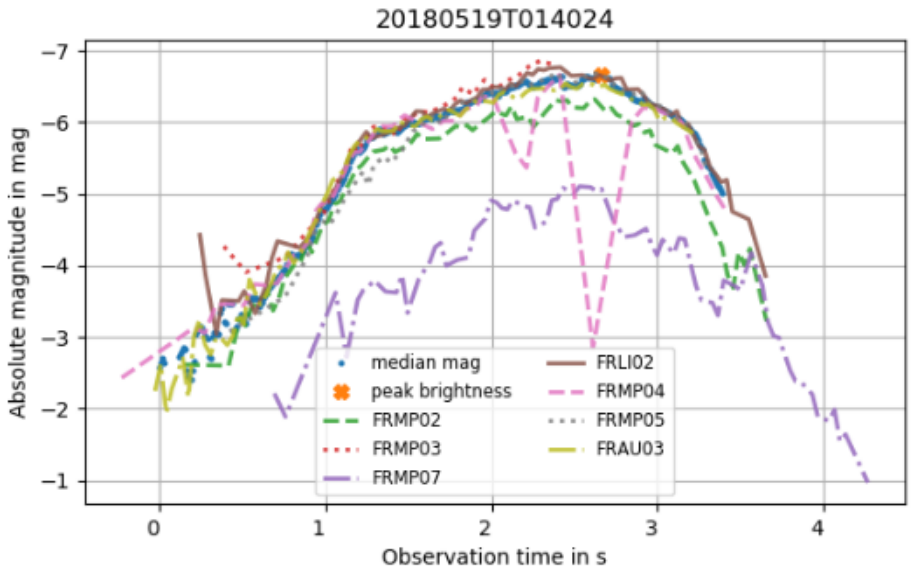

*Figure 2: Interpolated absolute magnitude data from seven single stations for the fireball from 2018 May 19 presented as color- and style-coded lines. The x-axis displays the time of observation of the fireball (in seconds), starting with the time of the first absolute magnitude value from the stations considered (first value of the computed median magnitudes). The median absolute magnitudes, calculated for each point in time the fireball was visible, are shown as blue dots. The maximum brightness of this median curve is highlighted with an orange 'x'.*

For the fireballs in the FRIPON database, the data are extracted and the brightness computed. Applying a least-square fit to the light curve the parameters $\mu$ and $\tau$ are found as explained in Section 2. In Fig. 3 the median light curve with resulting fit is presented for the fireball from 2018 May 19.

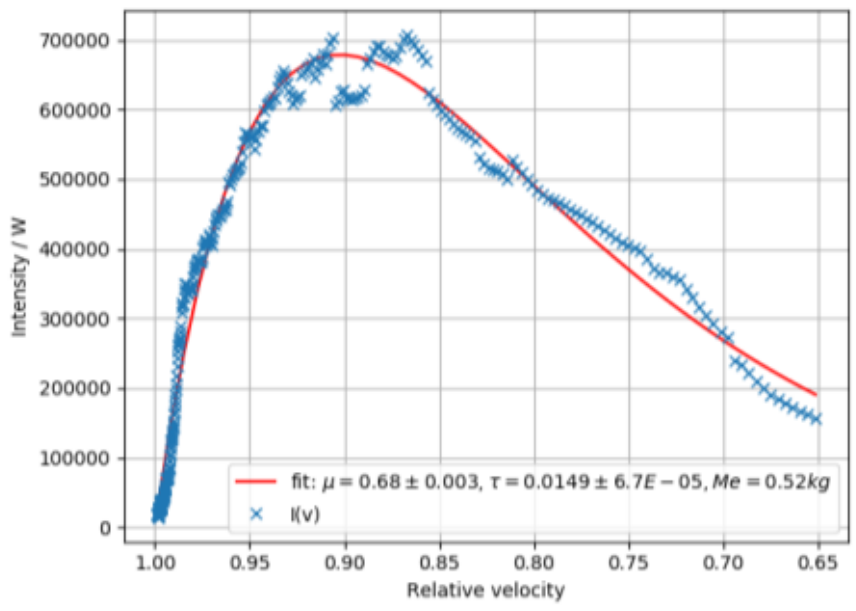

*Figure 3: The computed light curve of the fireball from 2018 May 19 with applied fit. The x-axis displays the relative velocity $v^*$ of the fireball for which each velocity value is divided by the fireball's initial velocity. The brightness values, calculated as explained above for each point in time for which a median magnitude value could be determined, are shown as blue 'x'. The applied fit is displayed as solid red line.*

## 6. Data and Results

In Colas et al. (2020) an overview of the FRIPON database and parameters of the recorded fireballs is presented. Status as of 2020 July 4, there are 3871 confirmed events in the database. For those events the parameters were extracted and the brightnesses and luminous efficiencies were computed as described in Section 5; see Table 2 for the fireball presented in detail in this work. In total, 1602 of the confirmed events had data sufficient to apply the method presented in this work. The other events were excluded from further calculations. The primary reason to exclude a fireball is the small number of cameras that detected the event. The method used in this work to calculate the light curve is only possible if at least two cameras have data from overlapping time ranges. This, as well as strong deviations in the individual cameras' magnitude data, led to insufficient brightness information for proper evaluations to be achieved. This situation often happens when significant weather differences are noticeable between the observing stations, mostly due to fog, clouds, etc, or when the fireball is only visible very close to the horizon in individual station's data.

Of the remaining fireballs, those that produced unrealistic results were also excluded. To do so, all fireballs with $0.01 > \mu > 0.73$ or $\tau > 1$ values were not included in further calculations. In addition, only those events with light curves allowing a qualitatively good fit were taken. To do so, the relative errors of $\mu$ and $\tau$ were computed based on the errors from the fitting process. Fireballs with a relative error for $\tau$ larger than 0.04, or with a relative error for $\mu$ larger than 0.1 were also excluded. Additionally, the remaining events were manually inspected and some with fits that did not match the curve well were also excluded. This mainly concerned fireballs for which only few data points were available. Hence, the determined error values of the fit parameters were relatively small and the events were not automatically excluded. This way a total of 294 fireballs with satisfactory parameters and manually confirmed good fits were left of the initial 3871 fireballs that were considered. Thus, less than 10 % of the whole available dataset was used for this work. It is mentioned here as we cannot rule out that this is introducing a bias in the study, for now.

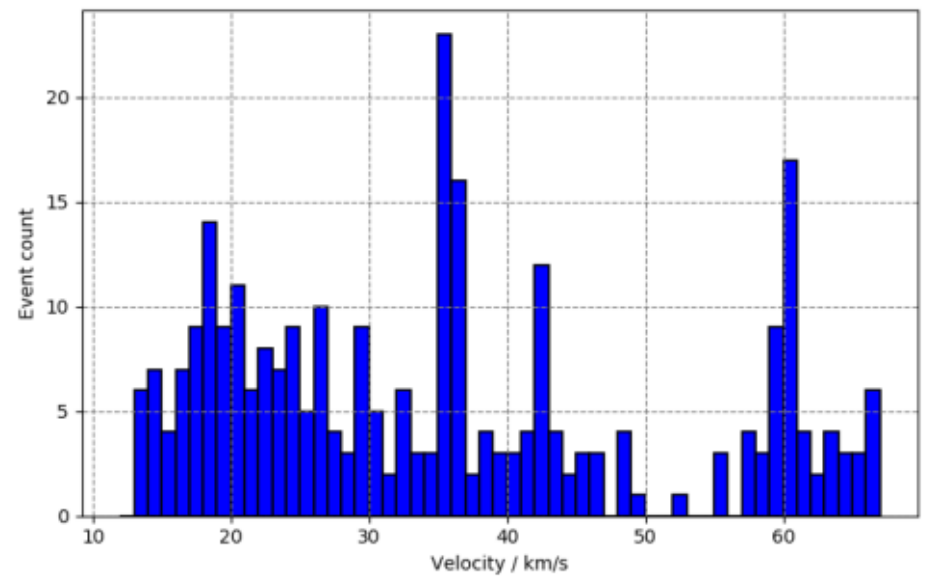

*Figure 4: Velocity distribution of the 294 FRIPON fireballs which meet the quality standards set for this study (see text for details) and form our subset.*

However, it should be mentioned that a bias could be present even within the entire sample. The investigation of possible observational biases should be part of future studies.

The velocity distribution of these 294 objects is shown in Fig. 4 and the mass distribution in Fig. 5. Fig. 4 indicates, that a large number of the analysed events are related to meteor showers. The peaks of the Perseids in August, with typical velocities of about 59 km s$^{-1}$ and of the Geminids in December, of about 35 km s$^{-1}$ (IMO 2017) are clearly visible. As mentioned in Colas et al. (2020) around 45 % of the recorded events in the FRIPON database are related to meteor showers.

The smallest object in our dataset has a pre-atmospheric mass of about $9 \cdot 10^{-6}$ kg, the largest one of ca. 32 kg. The median of the masses is $M_{e_median} = 0.012$ kg. As mentioned before, due to the different assumptions the mass values presented herein differ from the ones determined by the FRIPON pipeline. For the same density estimate, our values are about half an order of magnitude larger than theirs. The slowest fireball has a pre-atmospheric velocity of around 13.1 km s$^{-1}$, the fastest one of about 67.3 km s$^{-1}$. The median of the velocities is $v_{e_median} = 34.8$ km s$^{-1}$.

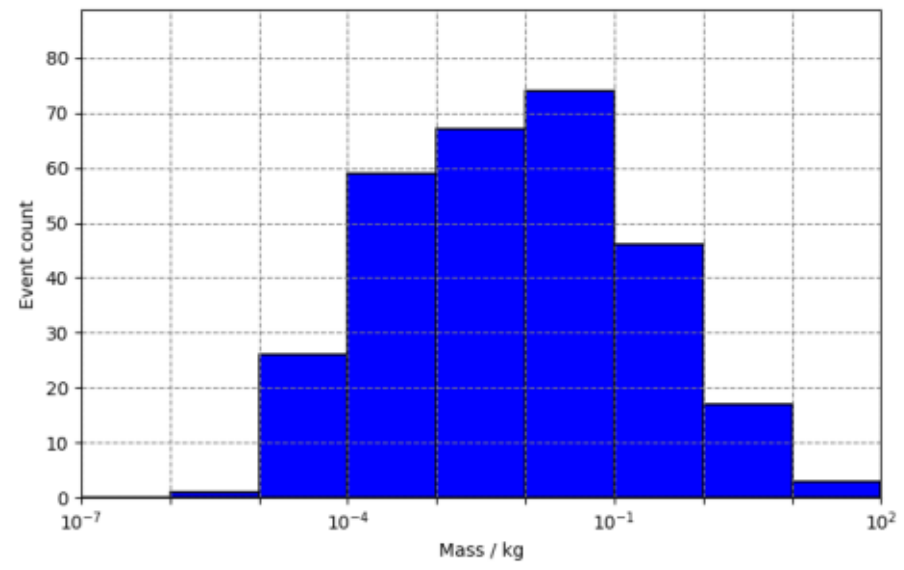

*Figure 5: Mass distribution of the 294 FRIPON fireballs which meet the quality standards set for this study (see text for details) and form our subset.*

Additionally, the distribution of the objects' entry angles is presented in Fig. 6. $\gamma$ describes the angle between the horizon and the trajectory. It is apparent that most meteoroids entered the Earth's atmosphere at an angle of about 45°, the median value is $\gamma_{median} =$ 39.3°. Since almost 2/3 of the events show $\gamma <$ 45°, the majority of them tend to impact at a rather shallow angle.

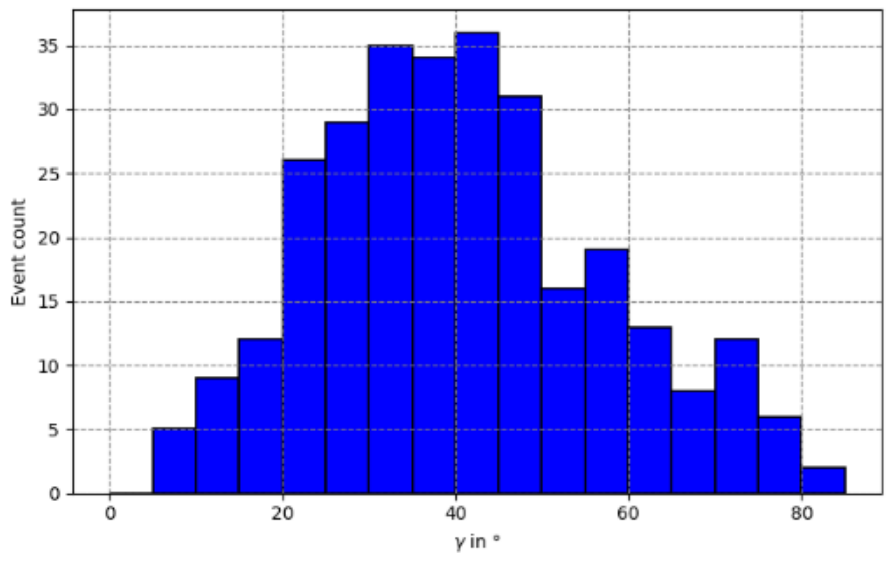

*Figure 6: Distribution of entry angles of the 294 FRIPON fireballs which meet the quality standards set for this study (see text for details) and form our subset.*

Fig. 7 shows the masses of the entering objects over their velocities. As expected, small objects were only detected if they were fast. This bias has to be kept in mind and should be further considered in future studies.

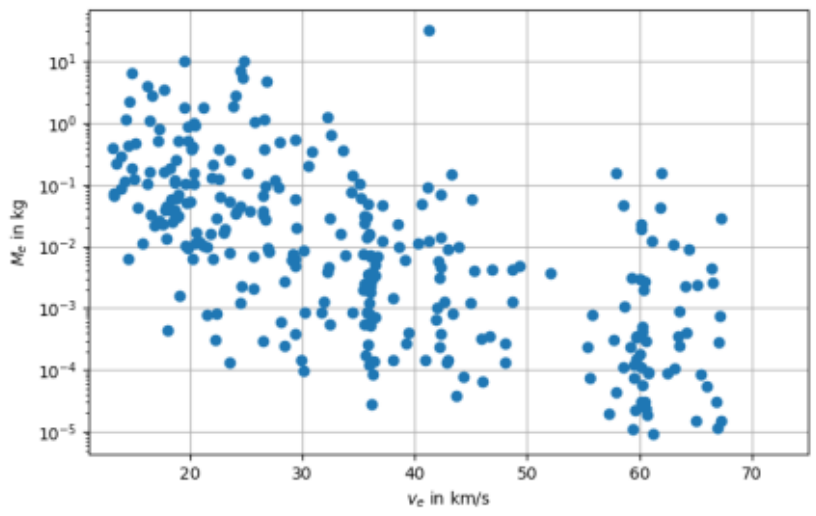

*Figure 7: Mass over velocity plot of the 294 FRIPON fireballs which meet the quality standards set for this study (see text for details) and form our subset..*

The distribution of the luminous efficiency $\tau$ is shown in Fig. 8, the distribution of the shape change coefficient $\mu$ in Fig. 9. As it can be seen our luminous efficiency values differ by orders of magnitudes ranging from $10^{-4}$ % up to 100 %. Ca. 70 % of the $\tau$-values are on the order of $0.1\ \% - 10\ \%$, only about 4 % are smaller (between $10^{-4}$ % and $10^{-1}$ %) and 26 % of the data showed a luminous efficiency in the range of 10 % – 100 %. The median of the $\tau$-values is $\tau_{median}$ = 2.17 %, the one of the shape change coefficient is $\mu_{median}$ = 0.61. Accordingly, if a single value to compute an object's mass from its brightness is required with no further information available, we would recommend a $\tau$ value of 2.17 %.

As it can be seen in Eq. (8) and (9) the shape change coefficient $\mu$ does significantly affect the form of the light curve. Hence, this parameter is another important topic, besides the question of luminous efficiency, to be addressed in this paper. A value of $\mu = 0$ corresponds to a case in which only the front of the object ablates but the cut-through surface area remains constant and only the front hemisphere changes. Hence, the maximal heating and evaporation occurs only in the vicinity of the front critical point of the meteoroid, the other parts of the body do not change during its way through the atmosphere. Many studies assume $\mu = 2/3$, like e.g. Halliday et al. (1996) or Jeanne et al. (2019). The case represents a uniform mass loss of the body over the entire surface area. Gritsevich & Koschny (2011) drew a comparison of the shape change coefficient to the rate of the meteoroid's rotation. According to them, a value of $\mu = 0$ corresponds to a case in which the entering objects shows a stabilized motion without rotation. In the case of $\mu = 2/3$ the rotation is rapid and chaotic allowing consideration of a nearly consistent body shape, shrinking uniformly. Eq. (8) indicates that the luminous efficiency has less influence on the shape of the light curve and rather influences the height of the intensity peak of the recorded curve, i.e. the intensity of the recorded fireball. Since $\tau$ indicates the percentage of initial kinetic energy that has been converted into light this is the expected behaviour. This could also point towards a bias in the observations. If only a small part of the energy is converted into visible light, the fireball might be too faint to be detected. Therefore, the distribution of the luminous efficiency presented in this work could underestimate smaller values of $\tau$.

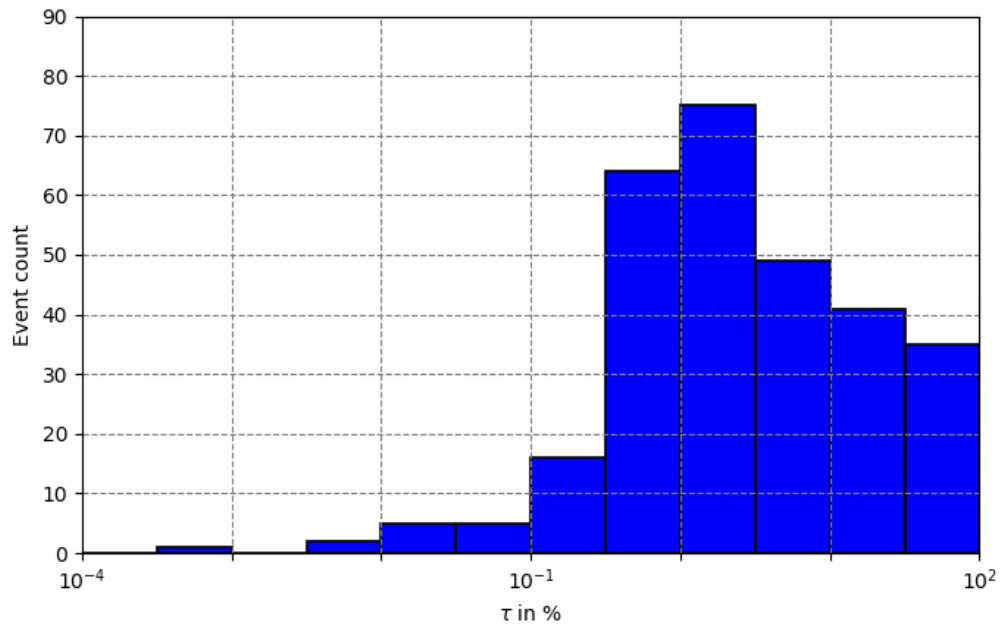

*Figure 8: The distribution of the luminous efficiencies τ derived in this work.*

The same is valid for further factors in Eq. (8), like the pre-atmospheric meteoroid mass and velocity $M_e$ and $v_e$, as well as the entry angle $\gamma$.

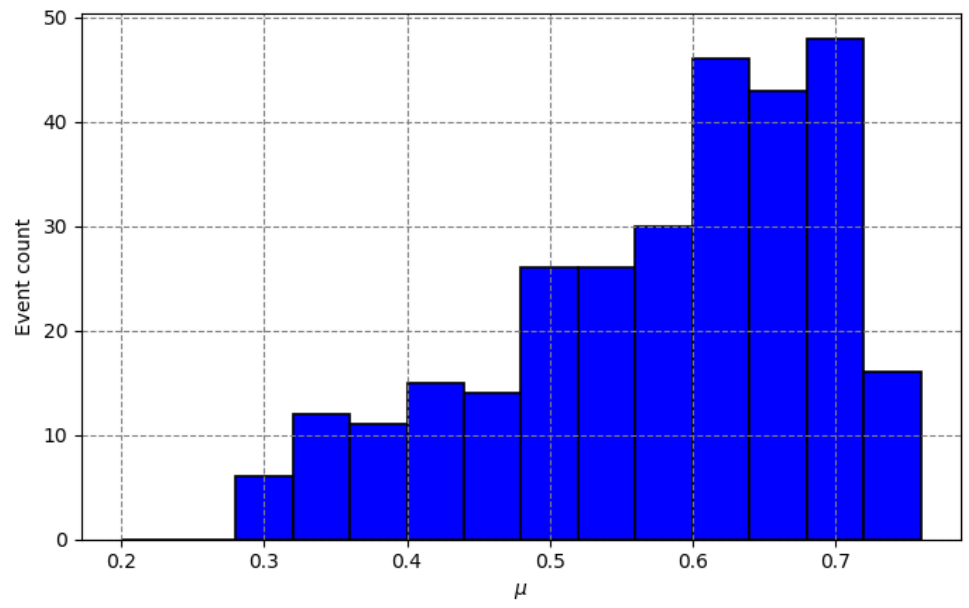

*Figure 9: The distribution of the shape change coefficient $\mu$ derived in this work.*

Since $\mu$ and $\tau$ represent physical properties of an event, they can be used to exclude unphysical objects from the analysis. Very small values of $\mu$ seemed to represent non-working fits to the light curve (which will be covered in more detail in the following). Hence values $\mu < 0.01$ were excluded. Moreover, based on the errors of $\mu$ and $\tau$ cases with fits that did not work well were excluded. These limit values were found by studying a large number of light curves and applied fits. Cases with relative errors of $\mu$ that are greater 0.1 or of $\tau$ larger than $0.04$ were excluded from the analysis. The relative error is also the basis for our upper limit of $\mu$. A value for $\mu$ of 2/3 should be seen as the physical upper limit. In combination to a relative error of less than 0.1 for this value, we accept values of up to $\mu$ = 0.73.

It is acknowledged that these thresholds were chosen based on subjective interpretations. An even more detailed study of the quality of the fits to the light curves and the light curves themselves could be considered in future work but goes beyond the scope of this study. The errors of $\mu$ and $\tau$ are based on the quality of the fit, too. Fig. 10 shows $\tau$ over $\mu$ with fitting errors for both values.

Furthermore, it should be kept in mind that the used model of the light curve does not take fragmentation into account. However, for some fireballs fragmentation is clearly visible in the light curves. Nevertheless, these events were included in this analysis as long as the fit was reasonably good. A more detailed analysis of the light curves and possible adjustment of the fit curves represents an interesting aspect for future works.

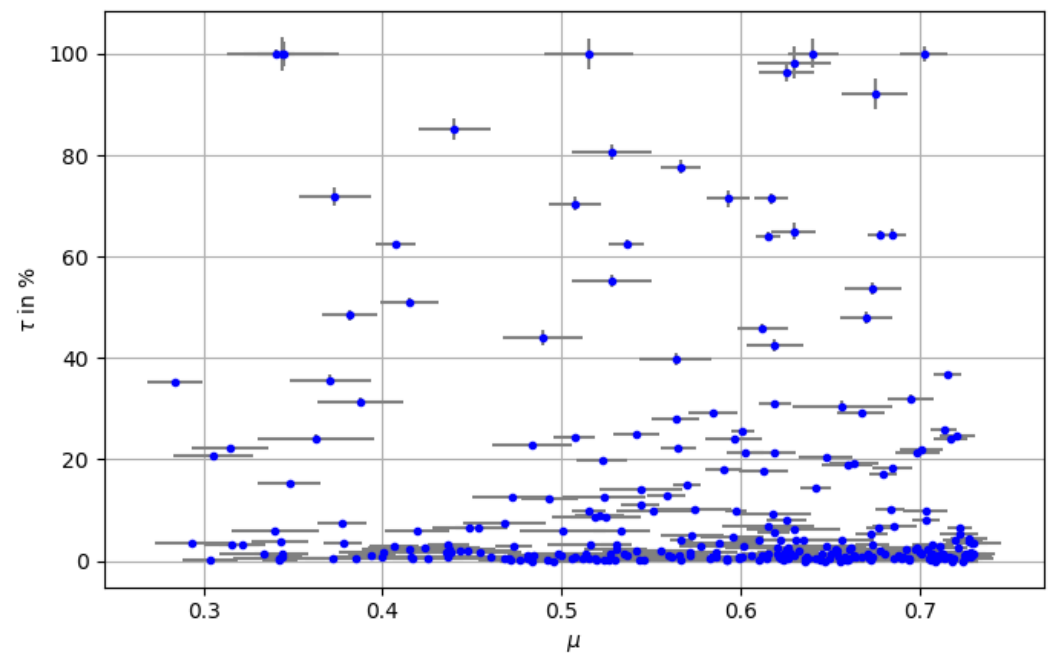

*Figure 10: $\tau$ over $\mu$ with fitting errors for both values.*

Another point to consider is the sudden increase in luminous efficiency as a consequence of abrupt meteoroid fragmentation. Such a behaviour can occur along the trajectory. It is usually observed in cometary aggregates that break apart when the loading hydrodynamic pressure reaches the disruptive strength of the particles. It was used by Trigo-Rodríguez & Llorca (2006, 2007) to infer the intrinsic strength differences of meteoroids from different parent bodies. The method was later applied during the preliminary examination team examination of 81P/Wild 2 to infer the strength of the particles penetrating in Stardust aerogel collector (Brownlee et al. 2006; Trigo-Rodríguez et al. 2008). The meteoroid break-up during the phase of heavy ablation causes the release of the small grains and the vaporization of volatile species that quickly impulse micron-sized dust to the shock front. As a consequence, large (mm to cm-sized) cometary aggregates often end into a bright flare that is far more luminous than the rest of the ionized column (Trigo-Rodríguez & Blum 2009; Trigo-Rodríguez et al. 2003, 2013). Obviously this effect should be considered to avoid luminous efficiency overestimation during bright ending flares.

The luminous efficiencies plotted over the fireballs' velocities are presented in Fig. 11, $\tau$ over the fireballs' corresponding meteoroids' masses in Fig. 12, and $\tau$ over the fireballs' entry angles $\gamma$ in Fig. 13.

In Fig. 11 a weak relation between the luminous efficiency and initial velocity of the fireball can be seen in semi-log space. $\tau$ seems to be larger for larger velocities. A relationship between the luminous efficiency and the initial meteoroid mass with negative linear behaviour can be clearly seen in log-log space in Fig. 12. Furthermore, a weak relation between $\tau$ and the entry angle is visible in Fig. 13.

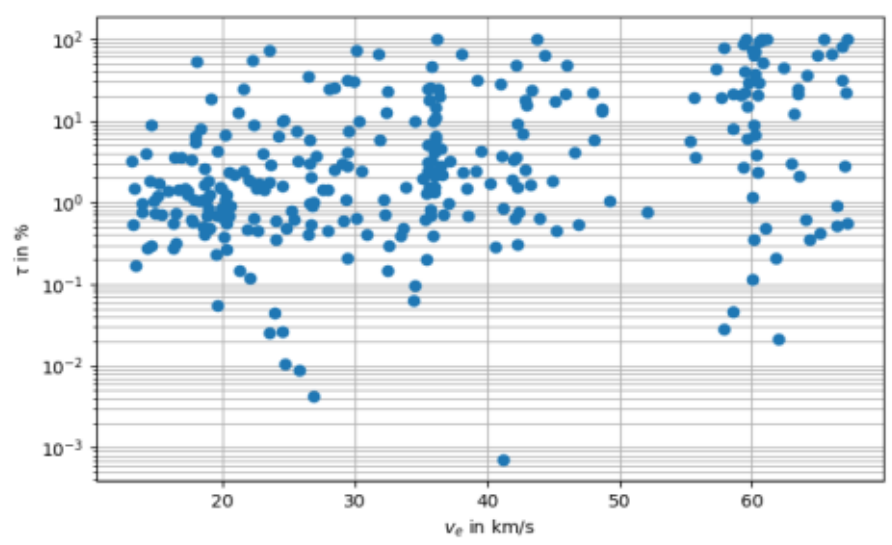

*Figure 11: The luminous efficiency τ over the fireball velocity.*

A possible dependence of $\mu$ on the mass, the velocity, and entry angle of the objects was also investigated, but no correlation was found. Furthermore, no dependency of either $\tau$ or $\mu$ on the peak brightness could be seen.

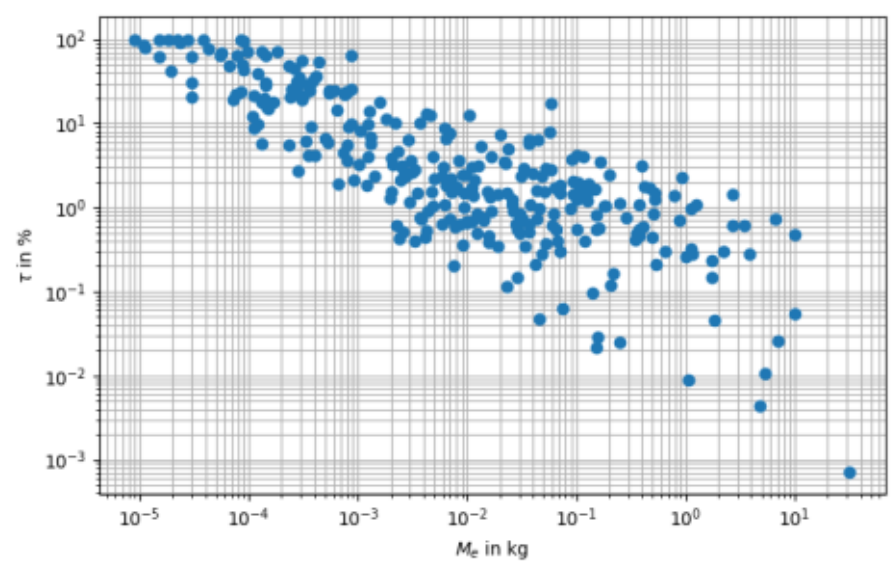

*Figure 12: The luminous efficiency τ over the fireball's corresponding entry object's mass in log-log space.*

## 7. Comparison of FRIPON data to literature

The objects analysed in the course of this work are in the mass range $10^{-6}$ kg – 100 kg. Our luminous efficiency values range from $10^{-4}\,\% - 100\,\%$, most are on the order of $0.1\,\% - 10\,\%$.

The luminous efficiencies presented in the Section 4 span a wide range of values. They are based on different types of studies of objects of different compositions in various velocity and size ranges. Our derived $\tau$-values are larger than those of most previous meteor and fireball studies dealing with smaller objects but consistent for studies of fireballs and larger asteroids. For the latter, the luminous efficiencies were also in the range of some 10 %.

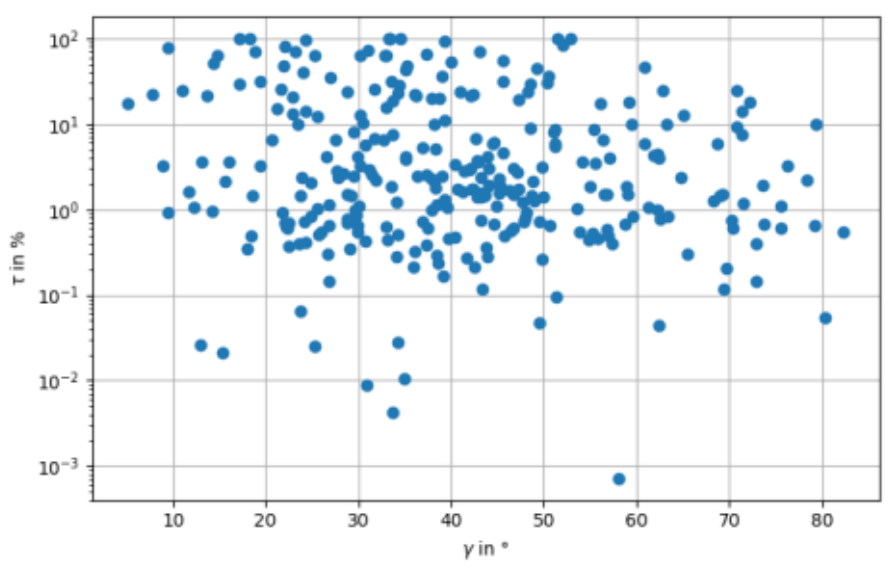

*Figure 13: The luminous efficiency τ over the fireball entry angle γ. The entry angle is measured from the horizon.*

Two examples of fireball studies were conducted by Halliday et al. (1996) and Ceplecha & McCrosky (1976). For the larger meteoroids, or even asteroids, their formulas might be more valid. Their values for the luminous efficiencies are on the order of some percent and both found a velocity dependency of $\tau$. Since the FRIPON data also consists mainly of fireballs, the results were expected to be most suitable for a comparison.

Halliday et al. (1996) derived the following relationships for $\tau$ (with $v$ in km s$^{-1}$)

$$\tau_{Halliday\ et\ al.} = 0.04\,, \quad v < 36 \text{ km s}^{-1}$$

$$\tau_{Halliday\ et\ al.} = 0.069 \cdot \left(\frac{36}{v}\right)^2, \; v \geq 36 \text{ km s}^{-1} \tag{14}$$

Ceplecha & McCrosky (1976) published a dependency of the luminous efficiency on the fireball velocity as presented in Eq. (15).

$$log(\tau_{\text{Ceplecha \& McCrosky}}) = -12.75 \quad \text{for } v < 9.3 \text{ km s}^{-1}$$

$$log(\tau_{\text{Ceplecha \& McCrosky}}) = -15.6 + 2.92 \cdot log(v) \text{ for } 9.3 \text{ km s}^{-1} < v < 12.5 \text{ km s}^{-1}$$

$$log(\tau_{\text{Ceplecha \& McCrosky}}) = -13.24 + 0.77 \cdot log(v) \text{ for } 12.5 \text{ km s}^{-1} < v < 17.0 \text{ km s}^{-1}$$

$$log(\tau_{\text{Ceplecha \& McCrosky}}) = -12.5 + 0.17 \cdot log(v) \text{ for } 17.0 \text{ km s}^{-1} < v < 27.0 \text{ km s}^{-1}$$

$$log(\tau_{\text{Ceplecha \& McCrosky}}) = -13.69 + log(v) \text{ for } 27.0 \text{ km s}^{-1} < v < 72.0 \text{ km s}^{-1} \quad (15)$$

Both were plotted in Fig. 14, taking the panchromatic response into account.

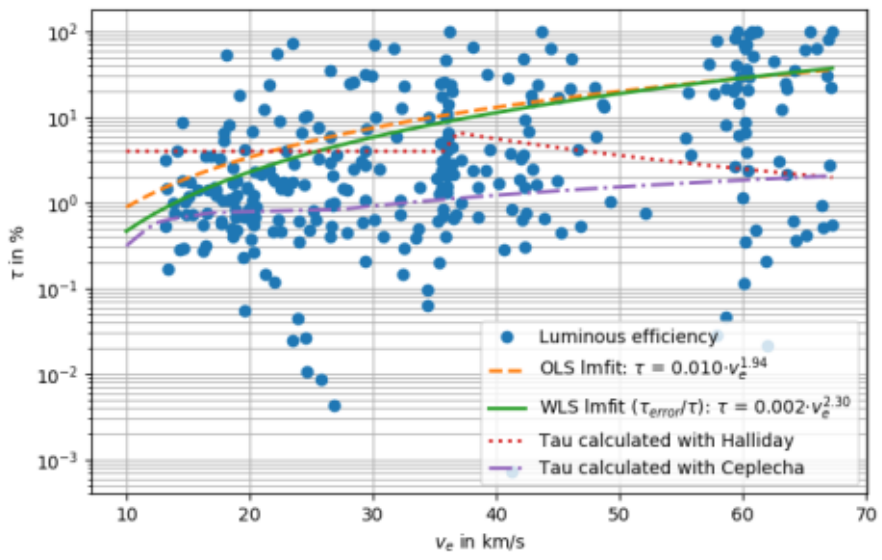

*Figure 14: Luminous efficiency over the initial velocity of the fireballs with fits through the data in semi-log space. Blue dots: values of $\tau$; orange dashed line: The ordinary least-square fit (OLS) applied to the data; green line: weighted least-square fit (WLS) applied to the data, the method weights the values by the relative error; red dotted line: the results found by Halliday et al. (1996); purple dashed-dotted line: the results found by Ceplecha & McCrosky (1976).*

To quantify the relation between the luminous efficiency and initial velocity of the fireball as seen in Fig. 11, a least-square fit was utilized to optimize the parameters of the function $\tau = b \cdot v_e^a$. This is presented in Fig. 14 in semi-log space. The ordinary least-square fit (OLS) gave $b = 0.010 \pm 0.013$ and $a = 1.94 \pm 0.31$, the weighted least-square (WLS) $b = 0.0023 \pm 0.0036$ and $a = 2.30 \pm 0.38$. The WLS method weights the values by the relative error of the luminous efficiency. Due to this, we expect the WLS method to represent the best estimate:

$$\tau = 0.0023 \cdot v_e^{2.3} \quad (16)$$

In this study, we also tested other fit functions and the distribution seems to show a quadratic dependence of $\tau$ on the velocity. Since the luminous efficiency represents a percentage of the object's energy, this relation seems to be reasonable.

However, as it can be seen the relative errors of the fitting parameters are quite large. This should be kept in mind as well as the large scattering the values exhibit, see Fig. 14. Yet, the luminous efficiencies are of comparable orders of magnitudes to the literature values.

In the most recently published studies by Subasinghe & Campbell-Brown (2018) and Capek et al. (2019), a weak relationship between the luminous efficiency and the initial meteoroid mass with negative linear behaviour in log-log space was found.

Converted to the formalism used in this work Subasinghe & Campbell-Brown (2018) found

$$\tau_{\text{Subasinghe \& Campbell-Brown}} = 0.0016 \cdot M_e^{-0.3647} \quad (17)$$

for masses in kg.

Capek et al. (2019) found a similar relation. Converted for masses in kg it would be expressed as:

$$\tau_{\text{Capek et al.}} = 0.01 \cdot M_e^{-0.38} \quad (18)$$

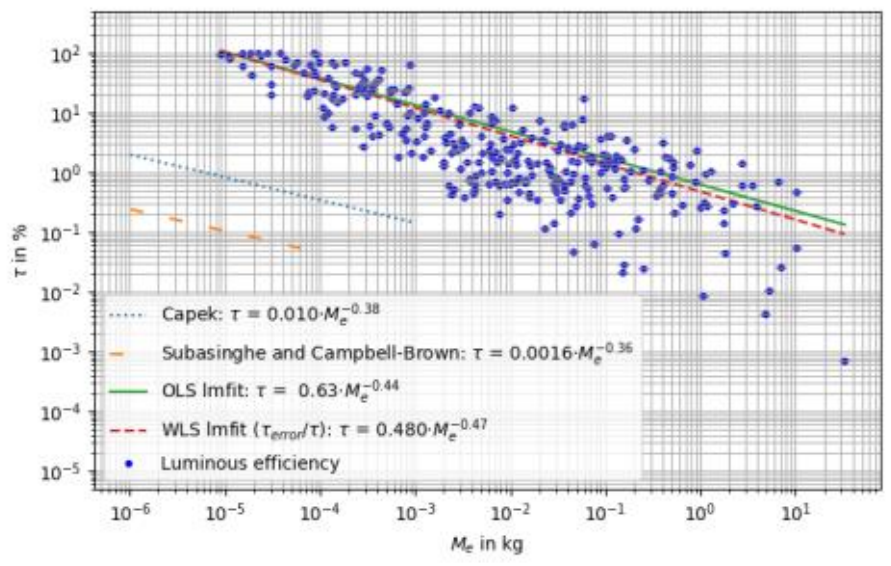

*Figure 15: Luminous efficiencies as derived in this work plotted over the fireball's corresponding meteoroid's mass with fit through the data in log-log space. Blue dots: values of $\tau$; green line: The ordinary least-square fit (OLS) applied to the data; red dashed line: weighted least-square fit (WLS) applied to the data, the method weights the values by the relative error; Blue dotted line: the results found by Capek et al. (2019); orange dashed-dotted line: the results found by Subasinghe & Campbell-Brown (2018).*

In Fig. 15 the luminous efficiencies are presented as they were derived in this work. They are plotted against the fireball mass with a fit through the data in log-log space. Following Capek et al. (2019) a fit with $\tau = \beta \cdot M_e^{\alpha}$ was applied to our data. The ordinary least-square fit (OLS) gave $\beta = 0.63 \pm 0.14$ and $\alpha = -0.44 \pm 0.02$, the weighted least-square (WLS) $\beta = 0.48 \pm 0.11$ and $\alpha = -0.47 \pm 0.02$.. Additionally, the results found by Subasinghe & Campbell-Brown (2018) and Capek et al. (2019) are included in Fig. 15 for comparison.

Again, we expect the WLS method to represent the best estimate:

$$\tau = 0.48 \cdot M_e^{-0.47} \tag{19}$$

In this work a similar slope of about $\alpha = -0.47$ was found. But, since our luminous efficiency values are larger, the value for $\beta$ shows a larger deviation. The luminous efficiency values derived in this study are about two orders of magnitudes larger than the values published in the works of Subasinghe & Campbell-Brown (2018) and of Capek et al. (2019) but show a similar dependency on the meteoroids' mass. It should be mentioned that in both works only 15, respectively 53, and especially smaller objects ($10^{-6}$ kg – $10^{-4}$ kg), were analysed. Nonetheless, the FRIPON data also include some meteoroids in the same size range.

The similar slope of all fits in combination with different y-axis intersections could point towards a systematic error in the brightness determination and efficiency assumptions. A systematic bias due to different heights of the investigated objects based on their different masses could also influence the results and is an interesting aspect for future studies, as we will show below.

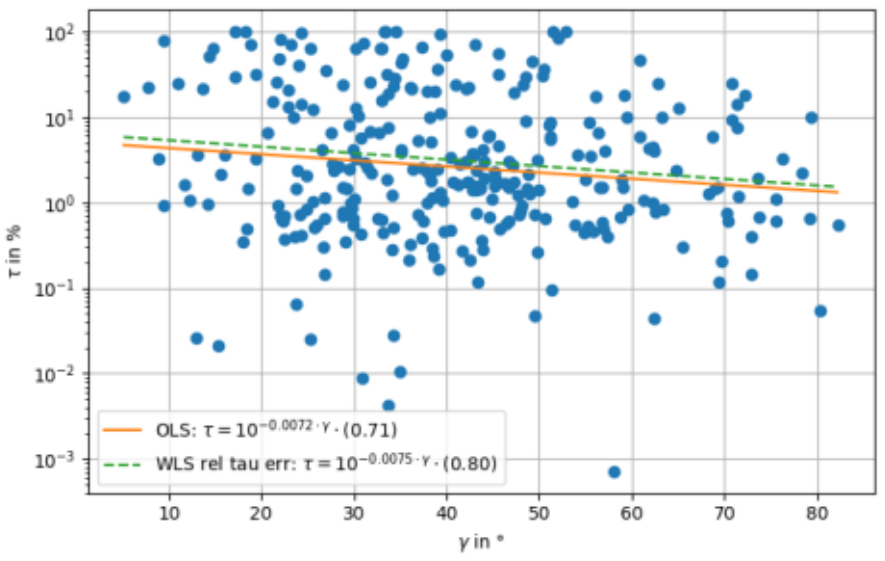

*Figure 16: Luminous efficiencies as derived in this work plotted over the fireball's entry angle with a fit through the data in semi-log space. Blue dots: values of τ; orange line: The ordinary least-square fit (OLS) applied to the data; green dashed line: weighted least-square fit (WLS) applied to the data, the method weights the values by the relative error.*

In Fig. 17 the dependence of $\tau$ on the analysed object parameters is visualized. The mass as well as the velocity of the corresponding object is presented in a 3D plot. $\gamma$ was omitted since the relation was only very weak.

For most of the fireballs, the luminous efficiency values computed in this work are around one order of magnitude larger than the $\tau$-values that are based on the photometric equations published in the literature. Depending on the results used for comparison, some luminous efficiencies differ by orders of magnitude. It has to be noted that the luminous efficiencies published in the literature, for which an overview is given in Section 4, also differ amongst themselves by orders of magnitude.

Interestingly, the studies on asteroids found rather larger values for $\tau$ in the range of some 10 %. However, our data indicate that the luminous efficiency decreases with increasing mass. The correlation between the angle of entry and $\tau$ was also based on studies on particularly large objects and the FRIPON data did only show a rather weak relation.

Nonetheless, to quantify the relation between the luminous efficiency and the entry angle of the fireballs a least-square fit was used to optimize the parameters of the function $\tau = \eta \cdot 10^{\zeta \cdot \gamma}$. This is presented in Fig. 16 in semi-log space. The ordinary least-square fit (OLS) gave $\zeta = -0.0072 \pm 0.0031$ and $= 0.71 \pm 0.14$, the weighted least-square (WLS) $\zeta = -0.0075 \pm 0.0032$ and $\eta = 0.80 \pm 0.14$. We expect the WLS method to represent the best estimate:

$$\tau = 0.80 \cdot 10^{-0.0075 \cdot \gamma} \tag{20}$$

Even if the slightly negative slope is in agreement with the work by Svetsov & Shuvalov (2018), since they found that especially objects with very shallow angles show large luminous efficiencies, we advise to not overestimate this weak link and to keep a rather large scattering in mind.

An important source of uncertainties are the investigated light curves themselves. So far, the brightness values are extracted from the single station data, averaged, and cleaned for outliers. It has to be discussed if the method is sufficient or if e.g. a station weighting by quality should be taken into account, similar to the approach Jeanne et al. (2019) carried out for the astrometric data reduction. Nonetheless, as examined before, we expect this method to be a good compromise between accuracy and computational effort due to the low-resolution of all-sky camera data.

Additionally, fragmentation was not taken into account during this study. The role of fragmentation was e.g. demonstrated in a study of the ablation of two very bright bolides (Borovička & Spurný 1996). As found later e.g. by Ceplecha & ReVelle (2005), this could introduce large errors in the derived values of the luminous efficiencies and shift them towards much higher values. We do expect that most of our analysed events fragment to at least some degree. Larger values of $\tau$ for fragmenting objects were also found e.g. by Subasinghe & Campbell-Brown (2018). For the investigated fragmenting meteoroids the luminous efficiencies increased from less than 1 % for most of their analysed meteors to $\tau$-values up to a few tens of percent.

Furthermore, the values which had to be assumed for the computations - the drag coefficient $c_d$, the pre-atmospheric shape factor of the meteoroid $A_e$, and the meteoroid bulk density $\rho$ - can affect the results significantly. As mentioned before, assumptions for these values vary a lot depending on the studies.

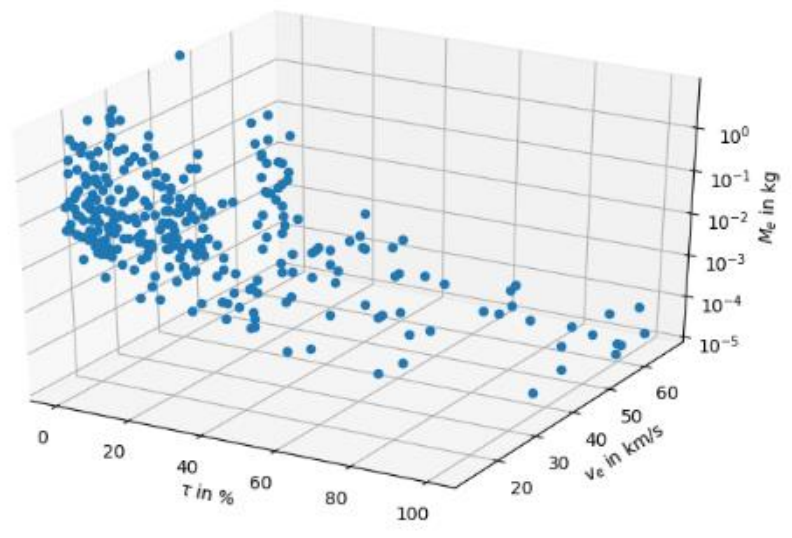

*Figure 17: The dependence of the luminous efficiency $\tau$ on the entry mass $M_e$ as well as on the entry velocity $v_e$ of the corresponding object presented in a 3D plot.*

Finally, the method itself should be investigated for possible error sources. Since the method is based on the analysis of deceleration data, it is expected that it is most robust for larger events with low end heights for which the deceleration is very prominent in the data. Fig. 18 presents the mass of the entering object over its end height, $h_{final}$. As expected, larger objects have considerably lower end heights.

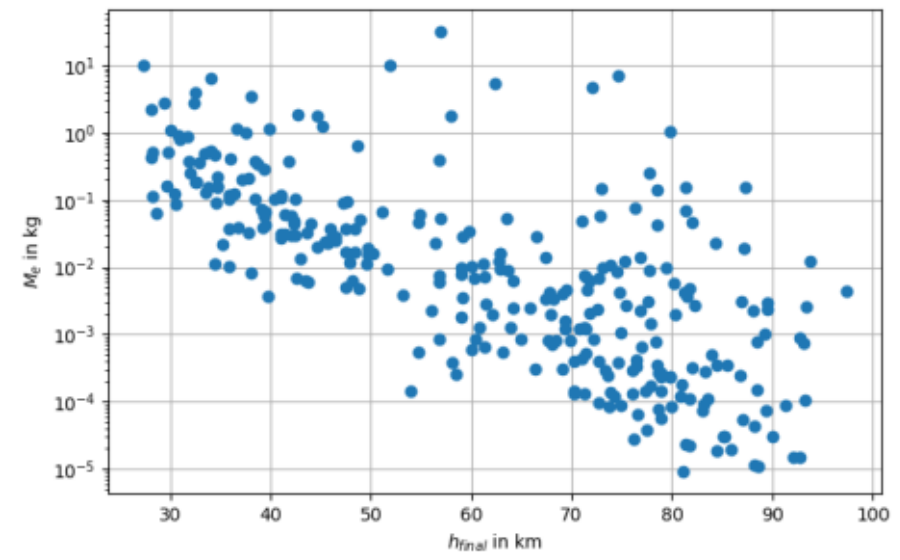

*Figure 18: The pre-atmospheric mass $M_e$ of the entering object over its end height $h_{final}$.*

The low number of small and fast events with larger end heights were detected by the system due to physical reasons not yet completely understood. According to the method used in this work, they would only be detected if almost all their energy were converted into visible light, i.e. their luminous efficiency would be around 100 %. This is the reason why an observational bias is expected and should be taken into consideration. Whether the used method is valid for those small objects at all has to be investigated further and in more details. Fig. 19 shows the distribution of $\tau$ over the pre-atmospheric mass of the object. It shows the same as Fig. 15 but only for the events with end height, $h_{final}$, lower than 55 km, representing a subset of 115 objects. It can be seen that the smallest meteoroids were excluded this way and that the luminous efficiency values in Fig. 19 are smaller than the ones in Fig. 15. This was expected since smaller objects tend to have higher end heights (compare Fig. 18) and the previously established trend towards larger $\tau$-values for smaller entry masses. The result of the WLS for this low end height subset is shown in Eq. (21). The fit has a slightly smaller shift but a steeper slope than the fit through all the investigated fireballs, compare Eq. (19).

$$\tau = 0.33 \cdot M_e^{-0.51} \tag{21}$$

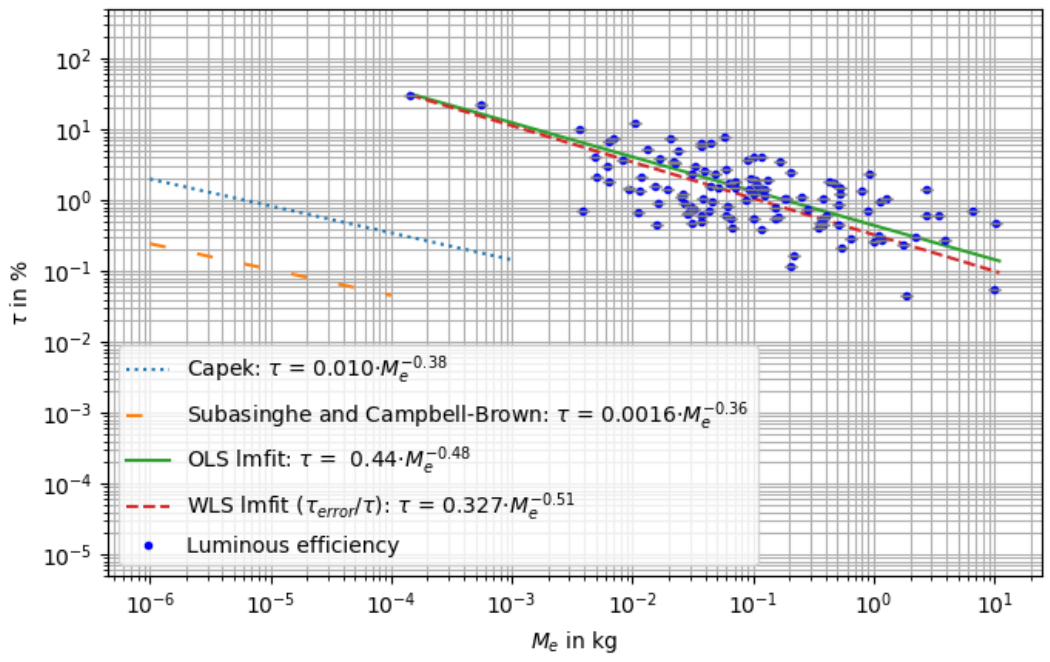

*Figure 19: Same as Fig. 15 but only for the 115 events with an end height lower than 55 km.*

For all events that were kept for the analysis, the final height at which it was last observed is calculated. Particularly low penetrating events with final heights lower than 55 km make up a subset of about 40 % of our complete dataset. Since larger meteoroids with higher tensile strength are able to penetrate the atmosphere deeper, most events with small initial masses and of fluffy cometary nature are not included in this subset. The degree of deepening is also associated with a higher tensile strength that informs us about the transition between fluffy cometary aggregates and rocky chondritic meteoroids (Blum et al. 2006, 2014; Trigo-Rodríguez & Blum 2009; Beitz et al. 2016). Additionally, we expect the events with low end heights and large masses to be the easiest ones to be observed with our method and thus, we assume a rather high quality of observations in those cases. We predict reliable $\tau$ calculations for these fireballs. However, we do not see that large initial mass events with final heights lower than 55 km produce lower luminous efficiencies, but rather some of those events with large masses and low $\tau$-values seem to end their flight in higher parts of the atmosphere. We find that this also impacts the dependency fit of $\tau$ on the initial mass which contrary to our expectations, was not shifted towards lower $\tau$-values. These discrepancies are further indications of an observation bias, as already discussed. This bias and possible indication of the limitations of the method used here should be further investigated.

## 8. Conclusion

One of the most important and controversially discussed meteor parameter for the conversion from magnitude to mass is the luminous efficiency $\tau$. In the course of this work the data of the FRIPON optical video camera system was used and a robust calculation method for the luminous efficiency was applied. We calculated the luminous efficiencies of fireballs using deceleration-based formulas and compared our results to values published in the literature. Luminous efficiencies and shape change coefficients were determined for a total of 294 fireballs with computed pre-atmospheric meteoroid masses in the range of $10^{-6}\ \mathrm{kg} - 10^{2}\ \mathrm{kg}$. We were able to confirm a dependency of $\tau$ on the velocity of the fireball in the investigated data. The relation $\tau = 0.0023 \cdot v_e^{2.3}$ was found. As expected, this implies that the radiation of faster meteoroids is more efficient. It could be explained as a consequence of higher excitation potentials and by the presence of a second spectral component of high temperature exhibiting important emission lines in the optical range. Moreover, a relationship between the luminous efficiency and the initial meteoroid mass with negative linear behaviour was found, $\tau = 0.48 \cdot M_e^{-0.47}$. This point to the fact that smaller meteoroids radiate more efficiently. The correlation between the angle of entry and $\tau$ did only show a weak dependency of: $\tau = 0.80 \cdot 10^{-0.0075 \cdot \gamma}$. Even if the luminous efficiency values derived in this work range from $10^{-4}\ \% - 100\ \%$ most of them are on the order of $0.1\ \% - 10\ \%$. This is considerably larger than in most meteor studies especially for smaller objects but well in the range found in studies of fireballs and larger asteroids. We recommend the median of the $\tau$-values $\tau_{median}$ = 2.17 % to be used if an independent luminous efficiency value is needed for fireball analysis and no additional information of the recorded objects are available. The dependency of $\tau$ on the meteoroid's mass with a slightly negative slope is in agreement with more recent studies.

A lot of interesting aspects to take into consideration for future studies are still open for interpretation and analysis. Future work might entail the study of more FRIPON data, using only the FRIPON data of highest quality, and/or taking into account the shower affiliation of the fireballs. Furthermore, to compare the obtained results with results based on data from other sensors or to apply another method to determine the luminous efficiency to the FRIPON data could be an interesting part of future works. Additionally, investigating the limitations of the deceleration-based method especially regarding the end heights of the recorded objects should be done in the next steps.

## 9. Acknowledgements

We thank the European Space Agency and the University of Oldenburg for funding this project. We would especially like to thank Maria Gritsevich for her support and constructive discussions on this matter. A special gratitude goes to the FRIPON Network for providing the data used for this study and to the FRIPON team for their support of this project.FRIPON was initiated by funding from ANR (grant N.13-BS05-0009-03), carried by the Paris Observatory, Muséum National d'Histoire Naturelle, Paris-Saclay University and Institut Pythéas (LAM-CEREGE). Vigie-Ciel was part of the 65 Millions d'Observateurs project, carried by the Muséum National d'Histoire Naturelle and funded by the French Investissements d'Avenir program. FRIPON data are hosted and processed at Institut Pythéas SIP (Service Informatique Pythéas), and a mirror is hosted at IMCCE (Institut de Mécanique Céleste et de Calcul des Éphémérides / Paris Observatory) with the help of IDOC8 (Integrated Data and Operation Center), supported by CNRS and CNES.
PRISMA is the Italian Network for Systematic surveillance of Meteors and Atmosphere. It is a collaboration initiated and coordinated by the Italian National Institute for Astrophysics (INAF) that counts members among research institutes, universities, associations and schools. The complete list of PRISMA members is available here: http://www.prisma.inaf.it. PRISMA was funded by 2016/0476 and 2019/0672 Research and Education grants from Fondazione Cassa di Risparmio di Torino and by a 2016 grant from Fondazione Agostino De Mari (Savona). FRIPON-Spain is coordinated from the Institute of Space Sciences (CSIC-IEEC). JMT-R, and EPA acknowledge financial support from the Spanish Ministry (PGC2018-097374-B-I00, PI: JMT-R).

## 10. Data availability

The data underlying this article were provided by FRIPON (Fireball Recovery and InterPlanetary Observation Network) by permission. Data will be shared on request to the corresponding author with permission of FRIPON.